%% file: ms.tex
\documentclass[twocolumn]{aastex631}

\usepackage{amsmath}
\usepackage{xspace}

\shorttitle{Gas Accretion From CGM at $z\sim2$}
\shortauthors{Vayner et al.}

\graphicspath{{./}{figures/}}

\received{2022 June 7}
\accepted{2022 November 29}

\submitjournal{MNRAS}

\begin{document}

\input{macros}

\title{Cold Mode Gas Accretion on Two Galaxy Groups at z$\sim$2}

\correspondingauthor{Andrey Vayner}
\email{avayner1@jhu.edu}

\author[0000-0002-0710-3729]{Andrey Vayner}
\affiliation{Department of Physics and Astronomy, Johns Hopkins University, Bloomberg Center, 3400 N. Charles St., Baltimore, MD 21218, USA}

\author[0000-0001-6100-6869]{Nadia L. Zakamska}
\affiliation{Department of Physics and Astronomy, Johns Hopkins University, Bloomberg Center, 3400 N. Charles St., Baltimore, MD 21218, USA}
\affiliation{Institute for Advanced Study, Einstein Dr., Princeton NJ 08540}

\author{Sanchit Sabhlok}
\affiliation{Department of Physics, University of California San Diego, 
9500 Gilman Drive La Jolla, CA 92093 USA}
\affiliation{Center for Astrophysics \& Space Sciences, University of California San Diego, 9500 Gilman Drive, La Jolla, CA 92093 USA}

\author[0000-0003-1034-8054]{Shelley A. Wright}
\affiliation{Department of Physics, University of California San Diego, 
9500 Gilman Drive, La Jolla, CA 92093 USA}
\affiliation{Center for Astrophysics \& Space Sciences, University of California San Diego, 9500 Gilman Drive, La Jolla, CA 92093 USA}

\author[0000-0003-3498-2973]{Lee Armus}
\affiliation{IPAC, California Institute of Technology, 1200 E. California Blvd., Pasadena, CA 91125}

\author{Norman Murray}
\affiliation{Canadian Institute for Theoretical Astrophysics, University of Toronto, 60 St. George Street, Toronto, ON M5S 3H8, Canada}
\affiliation{Canada Research Chair in Theoretical Astrophysics}

\author[0000-0002-6313-6808]{Gregory Walth}
\affiliation{IPAC, California Institute of Technology, 1200 E. California Blvd., Pasadena, CA 91125}

\author[0000-0001-7572-5231]{Yuzo Ishikawa}
\affiliation{Department of Physics and Astronomy, Johns Hopkins University, Bloomberg Center, 3400 N. Charles St., Baltimore, MD 21218, USA}

\begin{abstract}

We present Keck Cosmic Web Imager (KCWI) integral field spectroscopy (IFS) observations of rest-frame UV emission lines \ly, \civ $\lambda \lambda$ 1548 \AA, 1550\AA\ and \heii\ 1640 \AA\ observed in the circumgalactic medium (CGM) of two $z=2$ radio-loud quasar host galaxies. We detect extended emission on 80-90 kpc scale in \ly\ in both systems with \civ\, and \heii\ emission also detected out to 30-50 kpc. All emission lines show kinematics with a blue and redshifted gradient pattern consistent with velocities seen in massive dark matter halos and similar to kinematic patterns of inflowing gas seen in hydrodynamical simulations. Using the kinematics of both resolved \ly\ emission and absorption, we can confirm that both kinematic structures are associated with accretion. Combining the KCWI data with molecular gas observations with Atacama Large Millimeter/submillimeter Array (ALMA) and high spatial resolution of ionized gas with Keck OSIRIS, we find that both quasar host galaxies reside in proto-group environments at $z=2$. We estimate $1-6\times10^{10}$\msun\ of warm-ionized gas within 30-50 kpc from the quasar that is likely accreting onto the galaxy group. We estimate inflow rates of 60-200 \myr, within an order of magnitude of the outflow rates in these systems. In the 4C 09.17 system, we detect narrow gas streams associated with satellite galaxies, potentially reminiscent of ram-pressure stripping seen in local galaxy groups and clusters. We find that the quasar host galaxies reside in dynamically complex environments, with ongoing mergers, gas accretion, ISM stripping, and outflows likely playing an important role in shaping the assembly and evolution of massive galaxies at cosmic noon.

\end{abstract}


\section{Introduction} \label{sec:intro}

How massive galaxies form is one of the most puzzling questions in modern-day astrophysics. Distant ($z\sim 2$) quasar host galaxies with supermassive black hole masses of $10^{9}$ \msun\ are likely the progenitors of the most massive systems seen in the local Universe. These quasars reside in dark matter halos $\gtrsim 6\times10^{12}$ \msun \citep{myer06, tuml17, Geach19}. Such halos today have at most 10$\%$ of baryonic matter locked up in stars \citep{Kormendy13}. The majority of the baryonic matter resides in the circumgalactic medium (CGM) of these galaxies that extends to scales ten to a hundred times larger than the radius of the quasar host galaxies' stellar component \citep{tuml17,Silverman19,Zakamska19}. Furthermore, the majority of the metals that are the byproduct of stellar evolution reside in the CGM \citep{tuml17}. 

One major challenge in modern extragalactic astrophysics and cosmology is understanding the complex feedback loop between the supermassive black hole (especially in its actively accreting -- quasar -- phase), its host galaxy, and its CGM. For example, what is the role of gas in the CGM in fueling star formation and supermassive black hole growth? How do feedback processes inside a galaxy -- whether due to star formation or the supermassive black hole activity or both -- drive the metals and the gas back into the CGM? 

Simulations have indicated that accretion through cold streams ($10^{4}$K) is likely responsible for providing the majority of baryonic matter supply in distant ($z>2$) galaxies \citep{Keres09, vandeVoort11}. However, even though this matter is concentrated in filaments that trace the dark matter distribution, it has been observationally challenging to directly detect and study cold accretion streams in distant galaxies due to their highly diffuse nature. Historically, there have been two dominant ways of studying the CGM in distant galaxies. The first method is through transverse absorption line surveys (e.g., \citealt{zhu13, Prochaska14}) using background quasars or galaxies as continuum sources to probe absorbing CGM and direct ``down-the-barrel" spectroscopy of individual galaxies (e.g., \citealt{Steidel10, Rubin12}). The second method to study the CGM is using direct narrow-band imaging of \ly\ emission of high-redshift galaxies \citep{Cantalupo17}.

One of the best means of studying the resolved CGM around high redshift quasars is through \ly\ emission, as it is thought to be the brightest observable emission line in the CGM of active galaxies. The exact balance of energy powering \ly\ emission in CGM remains uncertain \citep{treb16}. The different energy sources include shock-heating of CGM by outflows (e.g., \citealt{tani00}), the release of gravitational binding energy during gravitational collapse of the gas onto the halo (e.g., \citealt{haim00, Fardal2001}), and photo-ionization by the quasar itself (e.g., \citealt{cant05}) and by the cosmic ultraviolet background (e.g., \citealt{goul96}). The expected \ly\ emission from the CGM surrounding a luminous quasar is on the order of $(5-100)\times10^{-19}$\ferg$\rm arcsec^{-2}$, an order of magnitude greater than the photo-ionization by the cosmic ultraviolet background \citep{cant05}, making quasars ideal targets for current resolved CGM studies with modern-day integral-field-unit spectroscopy. However, other sources of ionization such as embedded star formation in satellite galaxies can also be contributing \citep{Mas-Ribas17a}. 

The recent advent of a large field of view optical integral field spectrographs like MUSE on VLT \citep{Bacon10}, and KCWI (Keck Cosmic Web Imager; \citealt{Martin10}) have opened a new window for studying the CGM in emission. These instruments allow mapping of the 2D distribution of CGM gas surrounding distant galaxies and quasar systems and measurements of the gas kinematics. Due to wavelength coverage, MUSE is primarily focused on studying the CGM of powerful AGN at redshifts $>$ 3. The blue sensitivity of KCWI allows for the first time to study the CGM of $z=2-3$ quasars through \ly\ emission using an 8-10 m class telescope. Around a typical quasar at $z=2-4$, both MUSE and KCWI have been able to detect extended \ly\ with 20 minutes to one hour on source exposure times \citep{Borisova16,Cai19,Arrigoni-Battaia19,OSullivan20,Travascio20}. Deeper observations can yield detections of fainter UV lines such as \heii\ and \civ, but there appears to be a dependence on the source population \citep{Cantalupo17}. Certain very deep observations sometimes only yield tentative detection of addition UV emission lines around luminous type-1 radio-quiet quasars \citep{Arrigoni-Battaia18, Cantalupo19} while others place very stringent limits and only detect additional UV lines by stacking multiple data sets \citep{Fossati21}.

Observations of luminous radio-loud quasars \citep{Heckman91b,Roche14,Shukla22} and high redshift radio-galaxies \citep{Villar-Martin03,Villar-Martin07,Humphrey07,Vernet17} show a much higher occurrence of extended \heii\ and \civ\ emission within their \ly\ halos even in relatively shallow observations. The dichotomy between the CGM of the radio-loud and quiet population is an ongoing area of research. Differences in environments, gas phase conditions and gas heating mechanisms may play a key role.

In this paper, we present KCWI observations of the warm ionized gas ($10^{4}$K) in the CGM of two quasar host galaxies. We target the rest-frame UV emission lines \ly, \civ $\lambda \lambda$ 1548 \AA, 1550\AA\ and \heii\ 1640 \AA. We present sample selection, summary of observations, data reduction, and emission line analysis in Section Section \ref{sec:obs}. We discuss individual objects in Section \ref{sec:indiv_obj}. We discuss the implication of the observed kinematics and dynamics of the CGM in Section \ref{sec:discussion}. We summarize our conclusions in Section \ref{sec:conc}. We use an $\rm H_{0}=67.8$ \kms\ Mpc$^{-1}$, $\Omega_{\rm m}=0.308$, $\Omega_{\Lambda}=0.692$ cosmology throughout this paper \citep{Planck13}. 

\section{Observations Data Reduction \& Analysis}\label{sec:obs}
In this section we outline the observations conducted as part of this study consisting of KCWI, Keck - OSIRIS LGS observations and ALMA.

\subsection{Sample selection}
\label{sec:sample_selection}
We present details of the initial sample selection in \cite{Vayner21c}. In short, we selected targets to be observable with the Keck I adaptive optics system with a redshift where the primary optical emission lines (such as H$\beta$, \oiii and \ha) are redshifted into good atmospheric windows in the near-infrared. The parent sample was further constrained to be radio-loud type-1 quasars with jets on galactic scales. All quasars were selected to have a quasar bolometric luminosity $>10^{46}$ \ergs\ and a radio luminosity at 178 MHz $>10^{44}$ \ergs. The general properties of our targets are presented in Table \ref{tab:gen_prop}. For detail on redshift measurement please see \citet{Vayner21c}, where they were measured from the spatially unresolved quasar narrow line region.

We followed up KCWI observations on a subset of objects at $z=2-2.4$ where the \ly\ and \heii\ emission lines fall within a single grating configuration of KCWI and do not overlap with any strong atmospheric emission lines. Object 4C 09.17 and object 7C 1354+2552 were selected for this article based on a wealth of multi-wavelength observations that can bridge the scales from the nuclear region of the quasar host galaxy to CGM scales. Both sources show evidence of outflows and are interesting laboratories to study the intricate balance between feeding and feedback at high redshift. In addition, sensitive ALMA observations are also available to trace the molecular gas of the quasar host galaxy and nearby satellite galaxies on similar spatial scale to the KCWI observations. Details on the ALMA observations and data reduction can be found in \citet{Vayner21d}. In short, we followed up targets from the parent sample with strong indication of ionized outflows on galaxy scales. We targeted the CO (3-2) or CO (4-3) molecular emission line to resolve and map the molecular ISM and detect evidence of feedback on the molecular reservoir or through detection of molecular outflows. The matching field of view of ALMA and KCWI also allowed us to search for companion galaxies through detection of the CO emission line within $\pm$ 2000\kms\ from the redshift of the quasar.

\begin{table*}[]\label{tab:sample}
    \centering
    \begin{tabular}{l|c|c|c|c}
         Name & R.A & Dec & z & L$_{\rm bol}$ [\ergs]  \\
         \hline
         4C 09.17 & 04:48:21.74 &+09:50:51.46 & 2.1083 & $2.88\times10^{46}$\\
         \hline
        7C 1354+2552 & 13:57:06.53 & +25:37:24.46 & 2.0320 & 2.75$\times10^{46}$\\
         \hline
         
    \end{tabular}
    \caption{General properties of the two targets part of this article.}
    \label{tab:gen_prop}
\end{table*}

\subsection{KCWI}
KCWI observations were obtained on the nights of November 22 and 23, 2017, October 2 and 3, 2018 and March 28, 2019. The KCWI observations were conducted with the medium slicer using the BL grating with the central wavelength of 4499 \AA\ with the KBlue filter. Our observations covered a wavelength range from 3500\AA\ to 5500\AA\ with a spectral resolving power R$\sim$ 1800 and a field of of view of 16.5\arcsec $\times$ 20.4\arcsec. The observations consisted of acquiring the quasar with the MAGIQ guider and centering the quasar in the KCWI field of view. We set the exposure time to 1200 seconds for quasar observations and 600s for dedicated empty sky observation for ideal sky subtraction. Sky observations were taken by offsetting the integral field unit (IFU) field of view 30-60\arcsec\ away from the extended \ly\ emission onto a previously selected empty sky region with no objects brighter than 24th magnitude in $R$-band selected from the Dark Energy Survey (DES). We followed the ABAAB observing sequence where A is the quasar observations, and B is the empty sky observation. We dithered each quasar observation using multiples of half-slicer steps perpendicular to the slices and random offsets of 1-3\arcsec\ in the parallel direction. A guider image was saved after each dither and sky offset movement to facilitate better absolute astrometry if necessary. We observed 4C 09.17 on source for a total of 3.0 hours (9$\times$1200s) and 7C 1354+2552 for 3.3 hours (10$\times$1200s).

\subsection{OSIRIS}
OSIRIS observations were taken in the laser-guide star (LGS) mode. Initial observations were taken with the Hn2 and Kn1 filters for 4C 0917 and in the Hn1 and Kn1 filters for 7C 1354+2552 using the 50 milli-arcsecond lenslet plate scale aimed to resolve and study the host galaxy of the quasar \citep{Vayner21b,Vayner21c}. Subsequent observations on October 15, 2019, were taken with the Hn2 and Kn1 filter for 4C 09.17 but using the 100 milli-arcsecond lenslet plate scale to achieve a larger field of view to study the more diffuse emission discovered with the initial observation and to help bridge the spatial scale gap between the initial OSIRIS observations and KCWI. Each OSIRIS observation began by acquiring the tip/tilt star and centering it in the respective science frame. We then offset to the quasar using a known offset from Gaia astrometry. Each OSIRIS observation consisted of 600s exposures, with a dedicated sky frame taken once an hour using small dithers between each science observation.

\label{sec:data_reduction_analysis}
\subsection{KCWI Data Reduction \& Analysis}
The data were reduced with the KCWI data reduction pipeline \citep{Morrissey18}, which performs bias subtraction, cosmic ray removal, scattered light subtraction, flat fielding, wavelength calibration, and spectra extraction into a three-dimensional data cube. For each 3D data cube, the pipeline also performs differential atmospheric refraction correction and flux calibration using observations of standard stars recommended by the KCWI instrument team. Sky subtraction was done by scaling the flux as a function of wavelength in the sky data cube to match the data. The flat fielding and wavelength calibration were done using observations of a white light source and ThAr/FeAr lamps at the start of each night. The spectra were extracted into a data cube with a native pixel size of the slicer IFU of 0.69$\times$0.29\arcsec. We construct a white light image for each data cube by taking an average along the spectral axis. Offsets between each cube due to dithering are calculated by cross-correlating the quasar's position in each data cube. Science observations taken at different sky position angles are rotated such that north is up based on the recorded World Coordinate System (WCS) in the fits header. Finally, we combine the cubes using the CWITools package \citep{CWItools}, re-sampling the cubes onto a common grid with a pixel scale of 0.3\arcsec$\times$0.3\arcsec. Using the arclines data we measure a line spread function (LSF) by isolating a single emission and fitting a Gaussian model in each spaxel. We measure a median LSF of 1.04 \AA\ based on the Gaussian model dispersion value, with a standard deviation of 0.043 \AA\ across the slicer field of view. These values correspond to a LSF of 83.93 $\pm$ 3.48 \kms\ across the field of view. We achieved a final sensitivity (2$\sigma$) at 4000 \AA\ over 2\arcsec $\times$ 2\arcsec\ area in a 10\AA\ window of 2.5$\times10^{-19}$ \ergs~cm$^{-2}$ arcsec$^{-2}$ and 3.7 $\times10^{-19}$ \ergs~cm$^{-2}$ arcsec$^{-2}$ for 7C 1354+2552 and 4C 09.17, respectively.

\subsection{OSIRIS Data Reduction \& Analysis}
Details of the OSIRIS data reduction can be found in \cite{Vayner21c}. In short, we used the OSIRIS data reduction pipeline version 4.1.0 \citep{OSIRIS_DRP,OSIRIS_DRP17,Lockhart19} that does standard near-infrared detector reduction steps, constructs three-dimensional data cubes, does scaled sky subtraction to remove the bright OH glow from the sky and mosaics observations at different dither positions into a single science data cube. 

\subsection{Point-spread-function subtraction}
For PSF subtraction, we followed the procedure outlined in \citep{Vayner16,Vayner21,Vayner21b}. In short, the PSF for each data cube gets constructed by using the wings of the broad emission lines (e.g., \ly, \civ) and the quasar continuum by isolating and averaging those data channels together. The constructed image is then normalized to the maximum flux and subtracted from the rest of the data cub while re-scaling to the maximum spatial flux at each data channel. We construct a white light image excluding emission line channels following the PSF subtraction. Spaxels that show flux with a standard deviation greater than three times the background value measured near the edge of IFU FOV have their spectra flagged and are excluded from further data analysis. These spaxels generally reside near the core of the PSF, within the full width at half maximum (FWHM) of the PSF, and are dominated by residuals from PSF subtraction. Other studies \citep{Inskip11,Borisova16} remove continuum emission before constructing the PSF. For our sources, the channel range used to construct the PSF image is selected such that the PSF dominates the resulting image.

\subsection{Astrometry alignment}
All of the multi-wavelength data sets used in this study have their astrometry aligned to the same world coordinate system using the quasar as the common source across all data sets. All coordinates are measured relative to the quasar position in each given data set. For optical and near-infrared data we use the centroid of the bright unresolved quasar continuum, while for radio data we use the spatially unresolved emission with the flattest radio spectrum \citep{Vayner21d}. For KCWI, the astrometric accuracy depends on how well we can find the centroid of the quasar in our white-light image. To compute the centroid, we fit a 2D Gaussian to the white-light image. We obtain an uncertainty on the centroid of 0.03\arcsec (RMS) for both data cubes, and we take this to be the relative astrometric error between the KCWI and the other multi-wavelength data sets. The centroid of the quasar in the OSIRIS, HST and ALMA data sets can be found to an accuracy better than 0.005\arcsec. Hence their relative astrometric errors are smaller compared to KCWI.

\subsection{Optimal emission line extraction \& moment maps.}

\begin{figure*}[!th]
    \centering
    \includegraphics[width=7.3in]{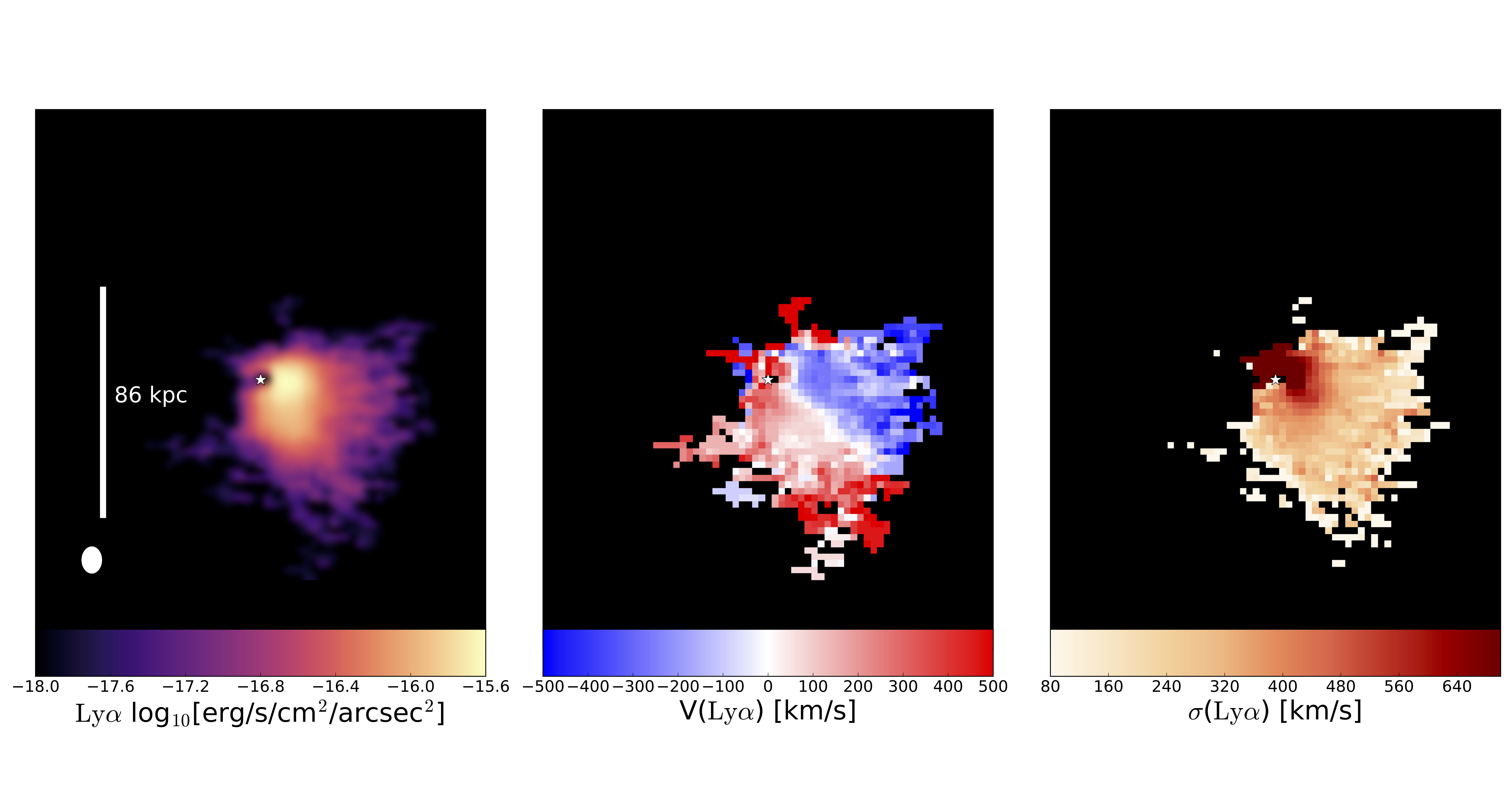}\\
    \includegraphics[width=7.3in]{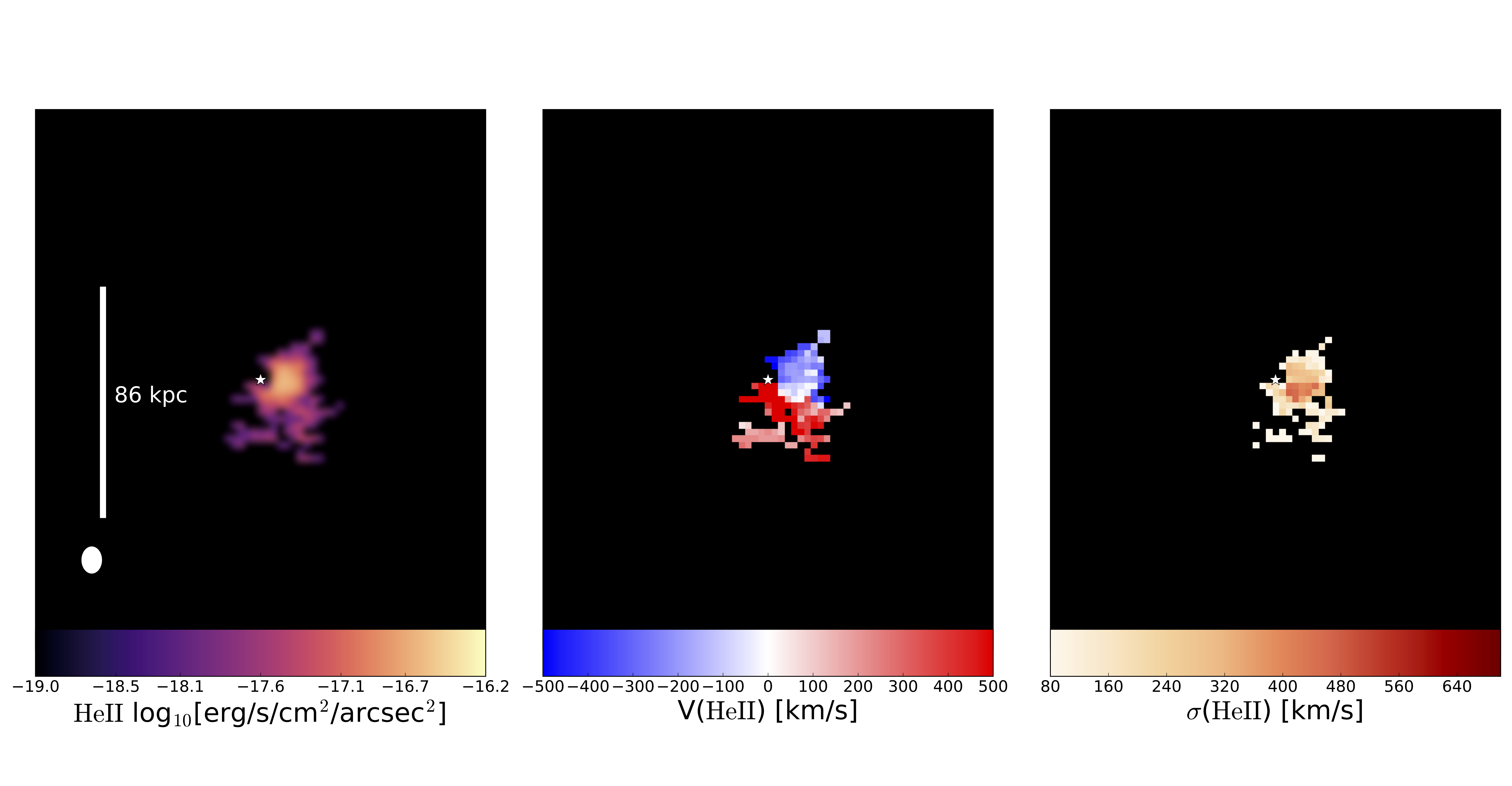}
    \caption{Moment maps for extended \ly\ and He II 1640 \AA\ emission in the 4C09.17 system. On the left, we present the optimally extracted extended line flux, the middle panel shows the radial velocity from the moment 1 map, and on the right, we show the velocity dispersion extracted from the moment 2 map. The bar in each right panel shows 10\arcsec\ or 86 kpc at the redshift of the source. The ellipse in the lower-left corner shows the FWHM of the PSF. North is up and east is to the left.}
    \label{fig:4C0917_LY_maps}
\end{figure*}

\begin{figure*}[!th]
    \centering
    \includegraphics[width=7.0in]{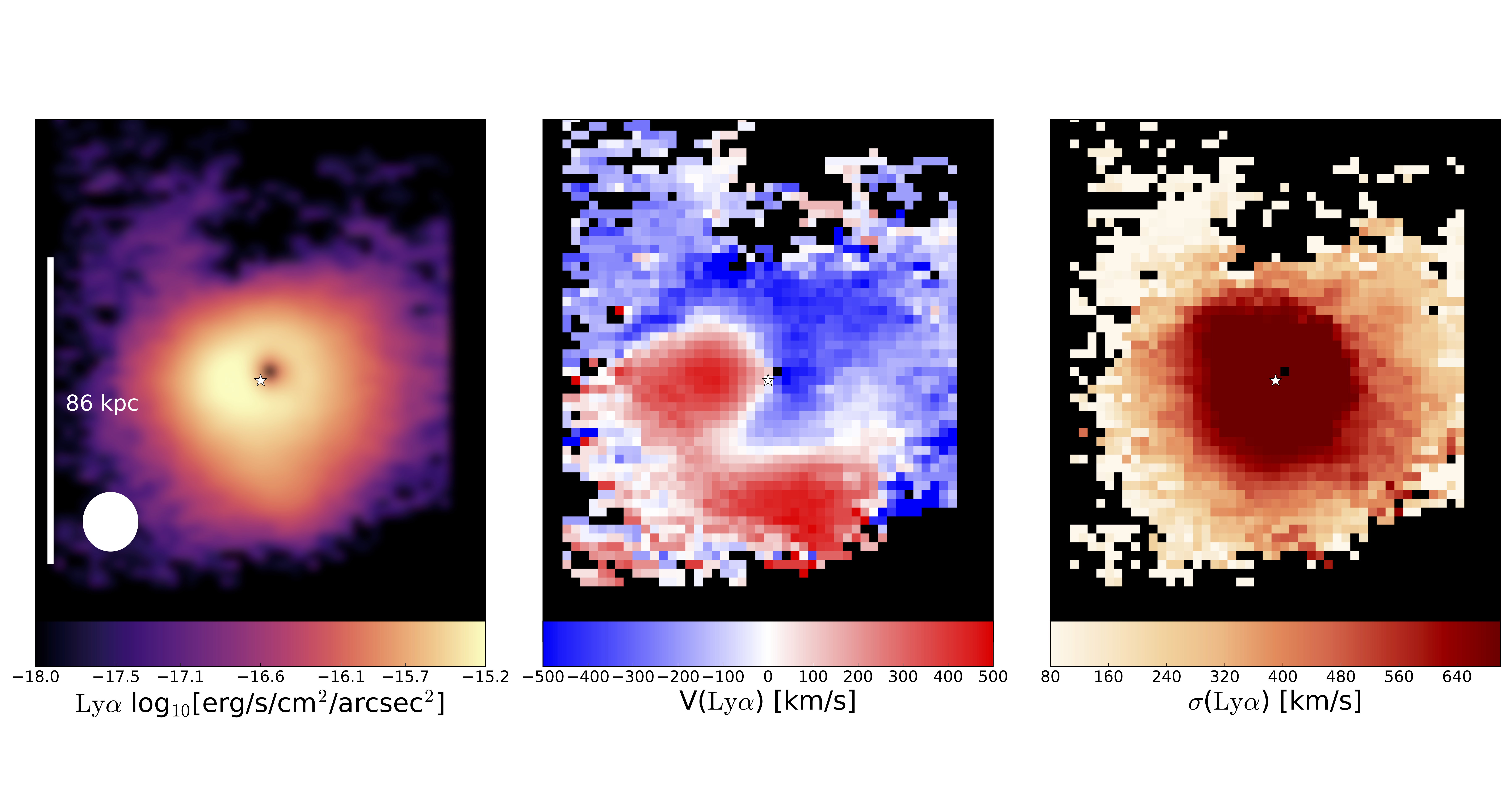}\\
    \includegraphics[width=7.0in]{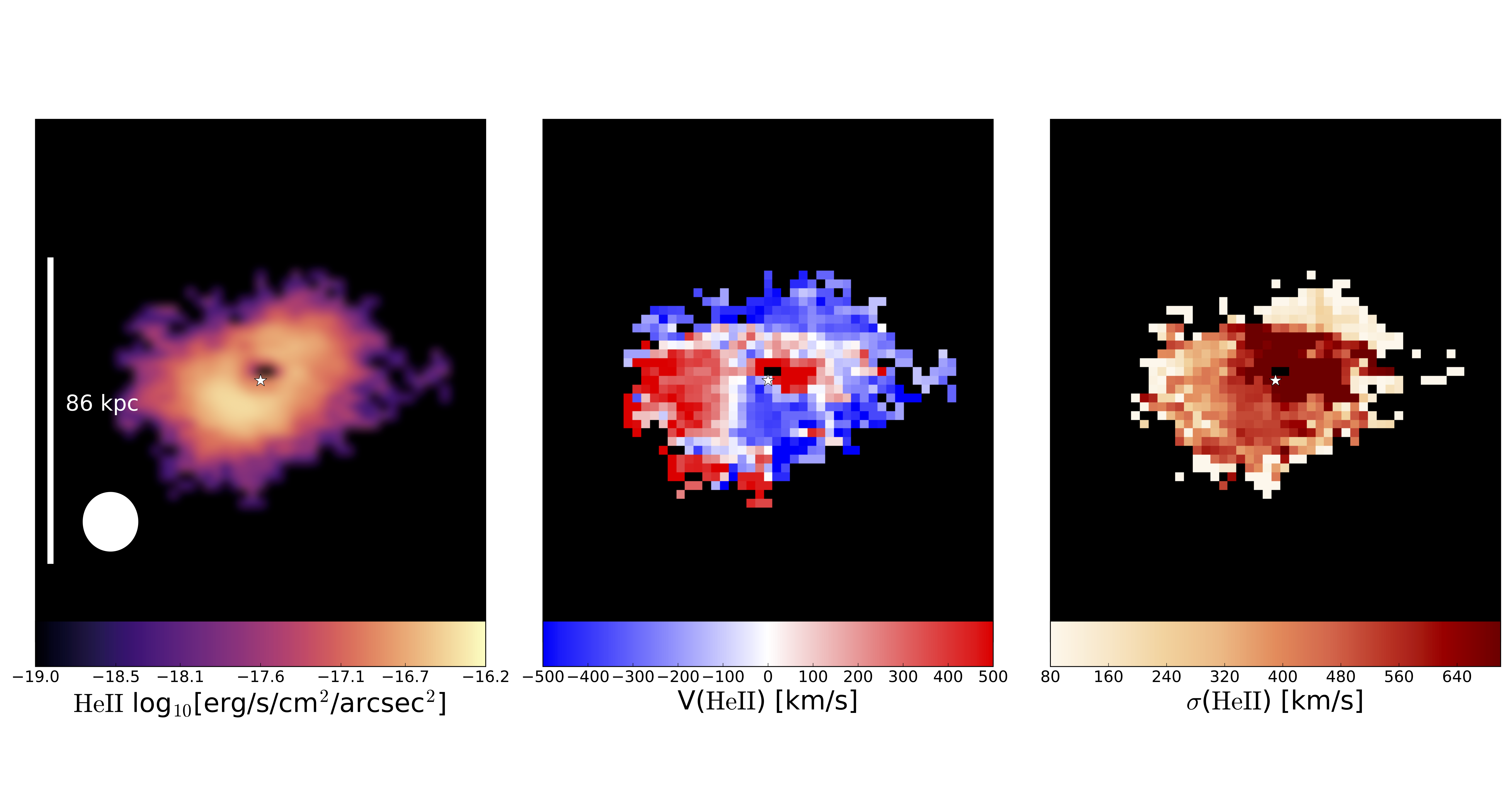}
    \caption{Moment maps for extended \ly\ and He II 1640 \AA\ emission in the 7C1354+2552 system. We present the optimally extracted extended line flux on the left, and the middle panel shows the radial velocity from the moment 1 map. We show the velocity dispersion extracted from the moment 2 map on the right. The bar in each right panel shows 10\arcsec\ or 86 kpc at the redshift of the source. The ellipse in the lower-left corner shows the FWHM of the PSF. North is up and east is to the left.}
    \label{fig:7C1354_LY_maps}
\end{figure*}

\begin{figure}[!th]
    \centering
    \includegraphics[width=3.5 in]{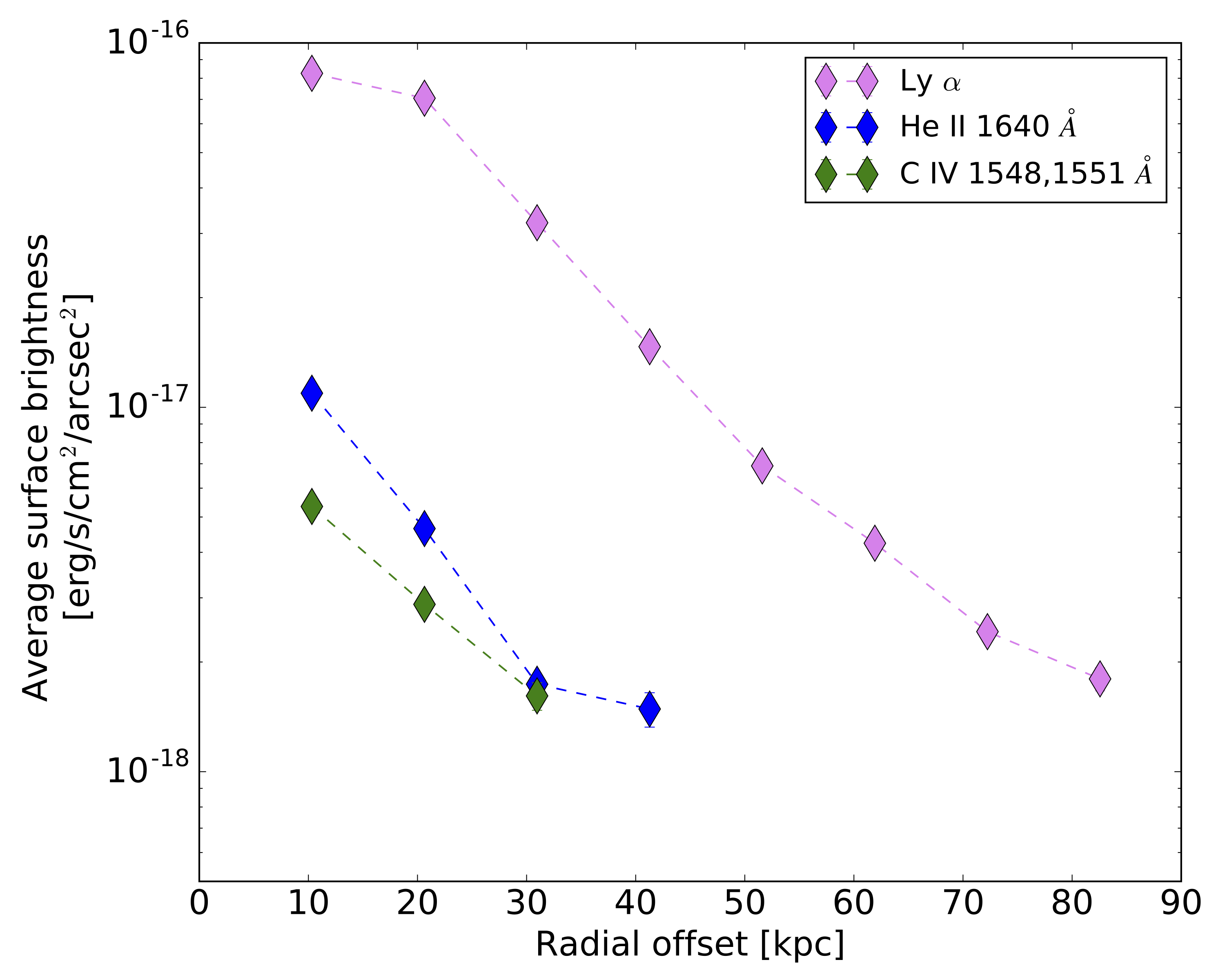}
    \includegraphics[width=3.5 in]{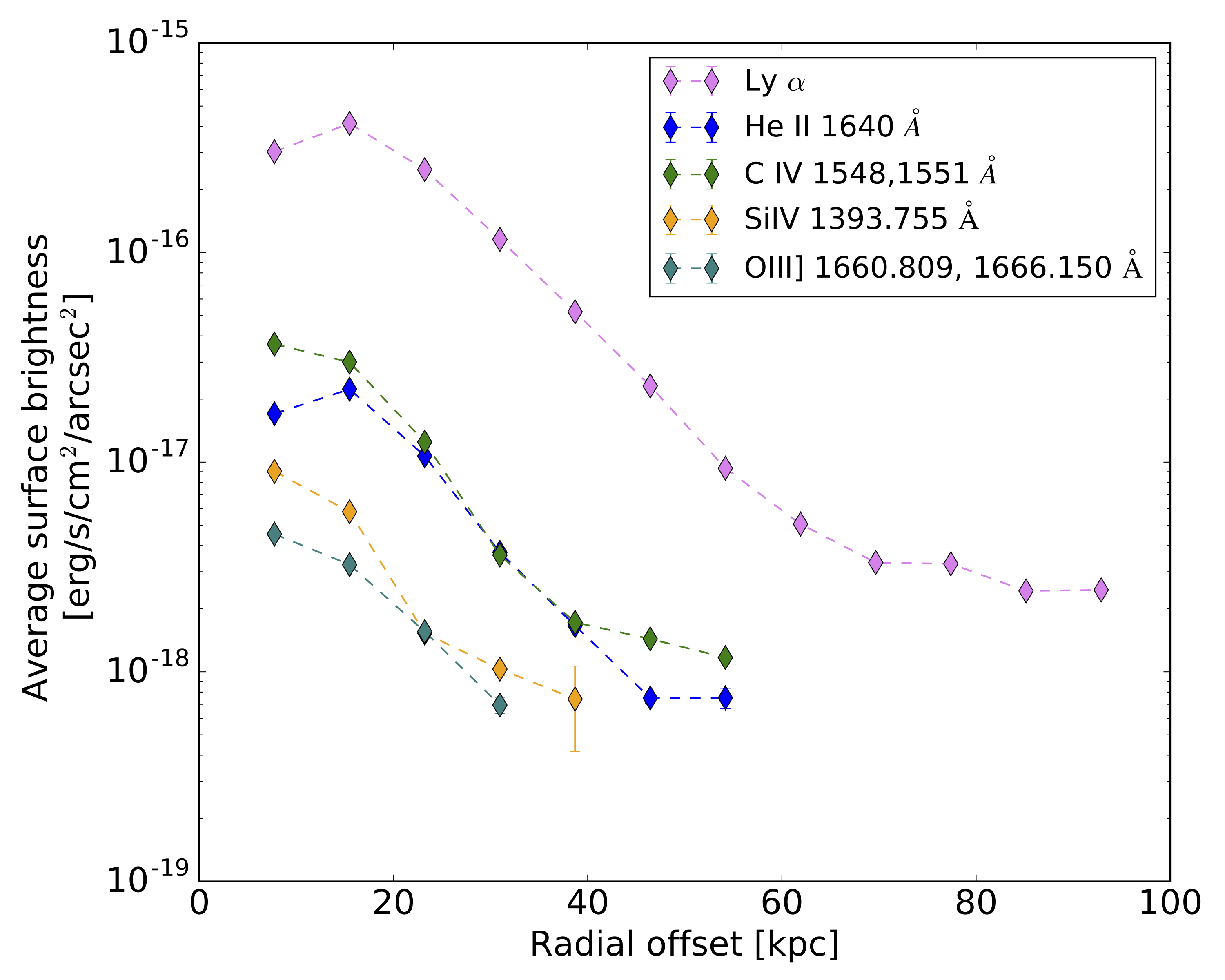}
    \caption{Average surface brightness profiles for the extended rest-frame UV emission line nebulae in the 4C 09.17 and 7C 1354+2552 systems.}
    \label{fig:SB_prof}
\end{figure}

We optimally extract the flux of each emission line using an ``object" data cube created through the use of segmentation maps. We select a minimum threshold of 2 $\sigma$ and use the \textit{cwi segment} routine within CWITools \citep{CWItools}, which loops through a range of wavelengths surrounding each emission line and creates a segmentation map based on the signal-to-noise ratio (SNR) criteria at each wavelength channel. The result is a data cube consisting of a mask at each wavelength location above the threshold. A moment-zero map in surface brightness units is created by summing the flux in each spaxel above the threshold criteria for each emission line. Moment 1 and 2 maps are created in a similar manner by applying the associated moment equation to flux in each spaxel above the threshold. We present moment maps for 4C 09.17 and 7C1354+2552 in Figure \ref{fig:4C0917_LY_maps} and \ref{fig:7C1354_LY_maps}. Moment 1 maps are created using the redshift derived from the \heii\ line. We construct radial surface brightness profiles from the moment 0 maps for each detected line out to the edge of the KCWI field of view to showcase the maximum detected spatial extent in each emission line (Figure \ref{fig:SB_prof}).

\section{Results: individual sources}
\label{sec:indiv_obj}
In this section, we discuss the results of each source by combining the multi-wavelength observations and the KCWI data to interpret the kinematics, dynamics, and photo-ionization condition of the gas on CGM scales.

\subsection{4C09.17}
\label{sec:4C09.17}

4C09.17 is a radio-loud quasar at z=2.1083, measured from the narrow line region of the quasar, with a bolometric luminosity of 2.88$\times10^{46}$\ergs. The quasar host galaxy consists of clumpy star-forming regions \citep{Vayner21b} with a star formation rate of 9$\pm$1 \myr\ and a molecular gas reservoir of $(3\pm0.3)\times10^{9}$ \msun \citep{Vayner21d}. We detect a multi-phase outflow in the quasar host galaxy extending in the eastern direction with a total outflow rate of 400$\pm$50 \myr\ likely driven by quasar activity \citep{Vayner21d}.

\begin{figure*}[!th]
    \centering
    \includegraphics[width=6.5 in]{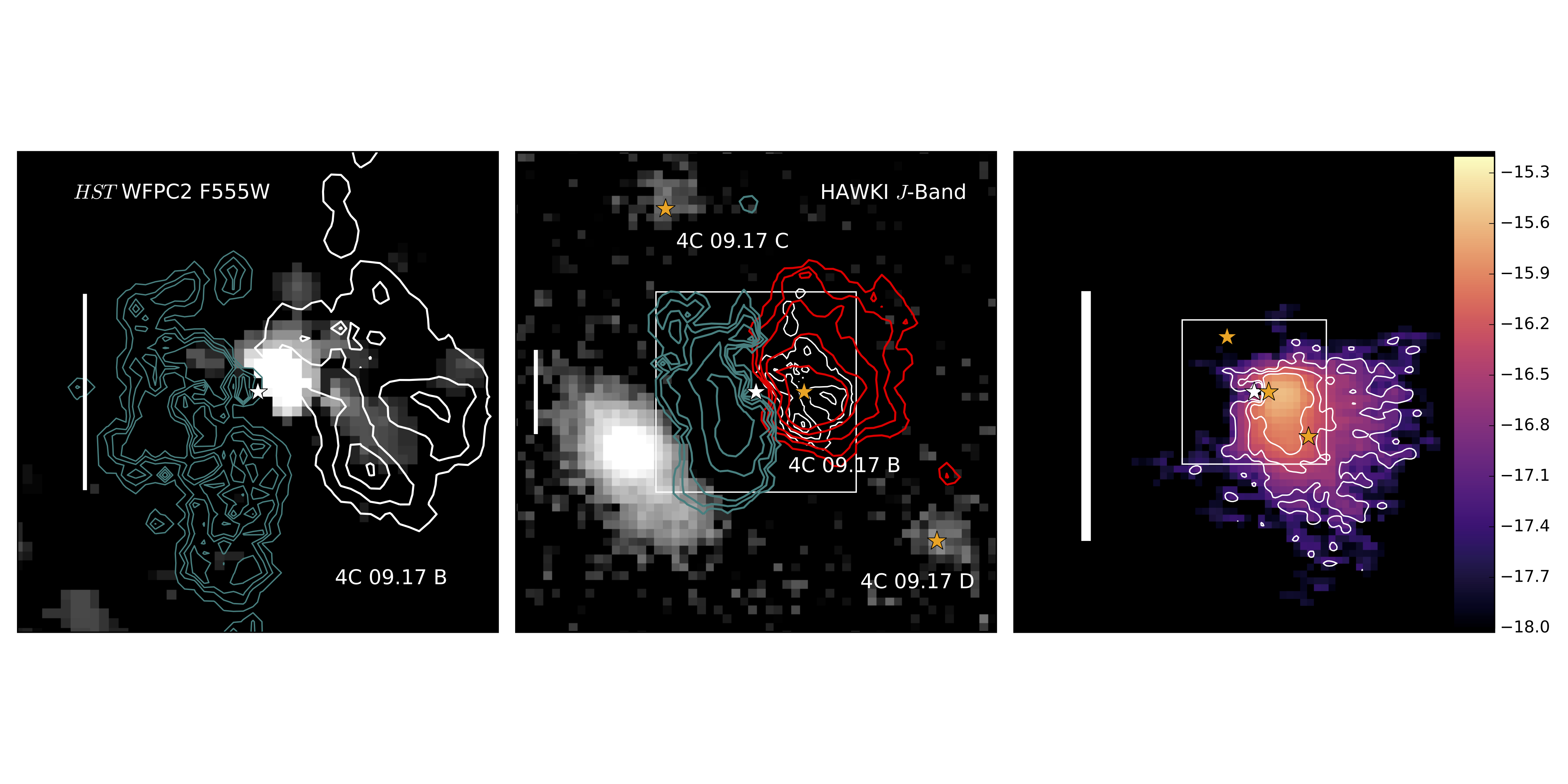}
    \caption{Detection of emission from stellar and gas components in 4C 09.17 on scales ranging from 1 to 90 kpc. The left panel shows the innermost emission detected with \textit{HST} WFPC2 imaging of rest-frame UV emission from massive young stars in the quasar host galaxy and in the nearby merging galaxy 4C 09.17 B. The white contours represent \ha\ emission in the quasar host galaxy and the nearby merger system originating from clumpy star-forming regions detected with OSIRIS-LGS observations. Teal contours show the location of the ionized outflow traced through \oiii\ emission. The middle panel shows near-infrared HAWKI $J$-band imaging of rest-frame $B$ band emission from stars in the host galaxies of the satellite galaxies C and D. Red contours in the middle panel show diffuse emission associated with the satellite galaxy B on scales of $\sim$ 16-20 kpc obtained with OSIRIS-LGS with a larger plate scale. Teal contours show the emission from the outflow, seen on larger more diffuse scales with the larger plate scale OSIRIS observations. The right panel shows extended \ly\ halo associated with the group system of galaxies. The orange stars represent the optical position of the galaxies, and the white star represents the location of the quasar host galaxy. We highlight the streams substructure in the \ly\ halo obtained by applying an unsharp-mask filter to the optimally extracted \ly\ emission map and present them as white contours on the right panel. North is up and east is to the left.}
    \label{fig:4C0917_system}
\end{figure*}

\citet{Lehnert92} detect potentially extended emission in the vicinity of the quasar. Higher resolution and more sensitive optical and near-infrared imaging reveal three galaxy candidates within 20 kpc from the quasar \citep{Armus97}. We confirm 4C 09.17 B \citep{Vayner21b} to be another galaxy merging with the quasar host galaxy through OSIRIS integral field spectroscopy by detecting emission from the \oiii, \ha, and \nii emission lines. The 4C 09.17 B galaxy consists of 4 star-forming clumps with a star formation rate of 96$\pm$8 \myr. We detect 4C 09.17 C through the CO (4-3) emission line in our ALMA program \citep{Vayner21d}. We also detect a molecular outflow in 4C09.17 C with an outflow rate of 2500 \myr\ in the southwest direction. We marginally detect a narrow \oiii\ emission line in the OSIRIS integral field spectroscopy of 4C 09.17 C; however, we do not detect an ionized outflow. We detect 4C 09.17 D part of the OSIRIS observations in the \oiii\ line obtained for this study that utilize the larger plate scale to detect more diffuse emission. The detection of 3 galaxies in the vicinity of the quasar host galaxy indicates that 4C 09.17 is a group system. While the definition of a galaxy group is ambiguous, the general consensus is that a galaxy group consists of several galaxies of similar mass with a combined total mass of $10^{13-14}$\msun. Given that the 4C 09.17 system consists of several galaxies with similar dynamical masses of $\sim10^{10-11}$\msun, their total mass including the dark matter halo can potentially surpass $10^{13}$\msun, hence, we believe this system to be a group. We present the position of the galaxies relative to the quasar that we detect both in $J-$ band continuum imaging and in ionized emission in Figure \ref{fig:4C0917_system}. We show the satellite galaxy properties from the multi-wavelength observations in Table \ref{tab:satellite_galaxies}. The discovery spectrum of each companion galaxy can be found in Appendix \ref{sec:appendix_sat}.

\begin{table*}[]
    \centering
    \begin{tabular}{c|c|c|c|c}
         Satellite galaxy & $\Delta$ RA & $\Delta$ DEC & Line & $\Delta$ V  \\
         \hline
         4C 09.17 B & 0.6\arcsec & 0\arcsec & \ha, \oiii, \nii  & -267$\pm$1 \kms \\
         4C 09.17 C & -1.13\arcsec & 2.29 \arcsec & CO (4-3), \oiii & 201$\pm$18 \kms \\
         4C 09.17 D & 2.26\arcsec & -1.86 \arcsec & \oiii & 411$\pm$20\kms\\
         \hline
         7C 1354 B & -0.82\arcsec & -0.11\arcsec & \oiii,\ha & 619$\pm$20 \kms \\
         7C 1354 C & -0.5\arcsec & 0.95\arcsec & CO (4-3) & -101$\pm$21\kms\\
         7C 1354 D & -3.75\arcsec &  3.5 \arcsec & CO (4-3) & -254$\pm$25 \kms\\
         7C 1354 E & 4.25\arcsec & 5.5\arcsec & CO (4-3) & -601$\pm$22 \kms
    \end{tabular}
    \caption{Properties of satellite galaxies in the group systems.$\Delta$ RA/DEC is the spatial offset relative to the quasar for each satellite galaxy. Line is the emission line used for spectroscopic redshift. $\Delta$ V is the relative offset of the satellite galaxy relative to the redshift of the CGM.}
    \label{tab:satellite_galaxies}
\end{table*}

With KCWI, we detect extended \ly\ emission with a maximum extent of 85.6 kpc, based on azimuthally averaged surface brightness profile (Figure \ref{fig:SB_prof}) extending towards the southeast. In addition, we detect \civ\ and \heii\ emission with a maximum extent of 31.5-40 kpc. The moment 1 map for each emission line features a velocity gradient with a direction of the maximal velocity gradient (major kinematic axis) along the north-west to southeast direction with a maximum velocity offset of $\pm$500 \kms. To correct for beam smearing in the central part of the CGM, we measure the velocity dispersion in the outskirts of the \heii\ and \ly\ nebulae and find a value of 230-307 \kms, away from the cusp of the velocity gradient. The measured velocity dispersion of \heii\ reflects bulk motions of the gas more accurately than that of \ly. 

\begin{figure}[!th]
    \centering
    \includegraphics[width=3.4 in]{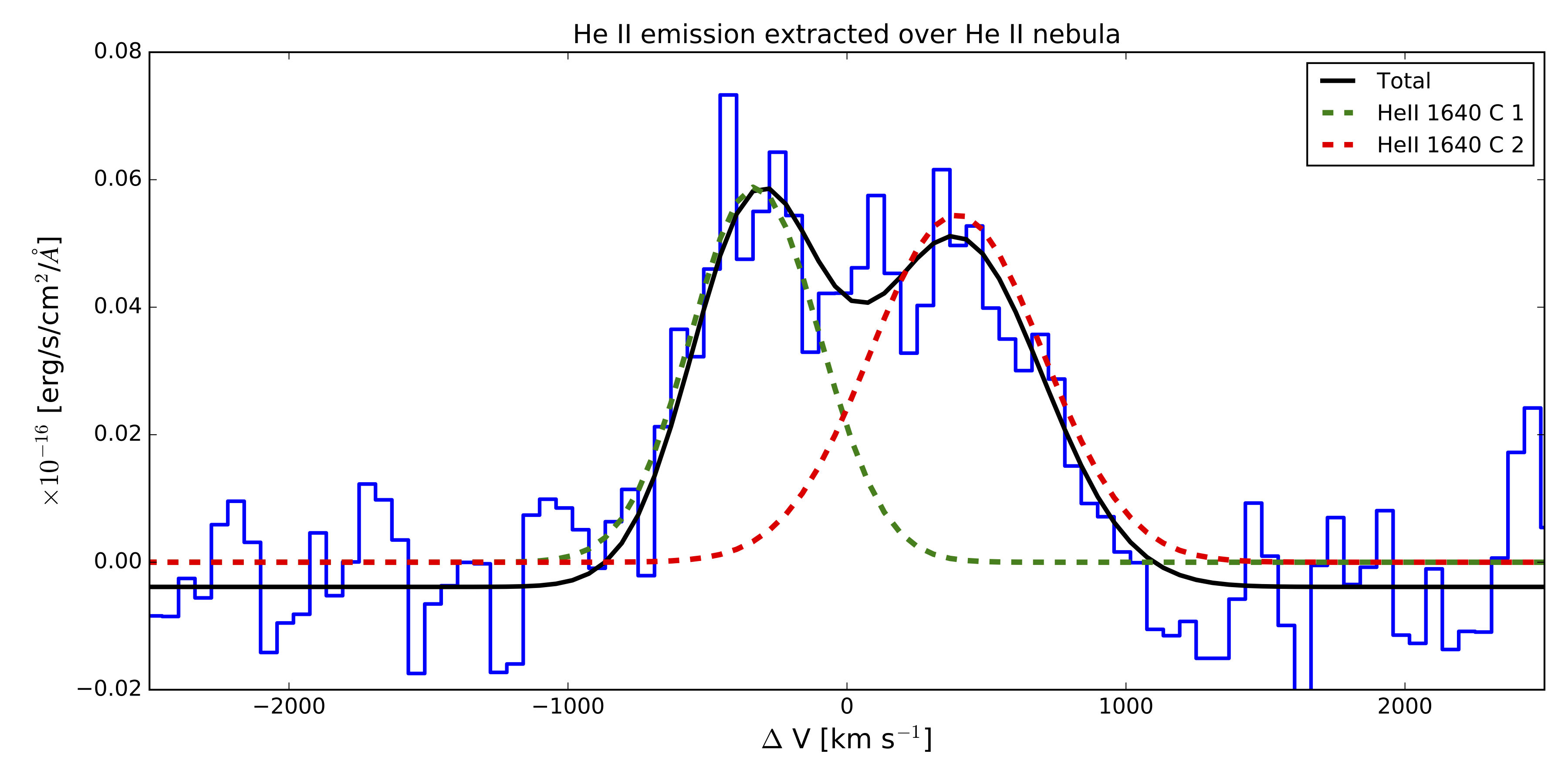}\\
    \includegraphics[width=3.5 in]{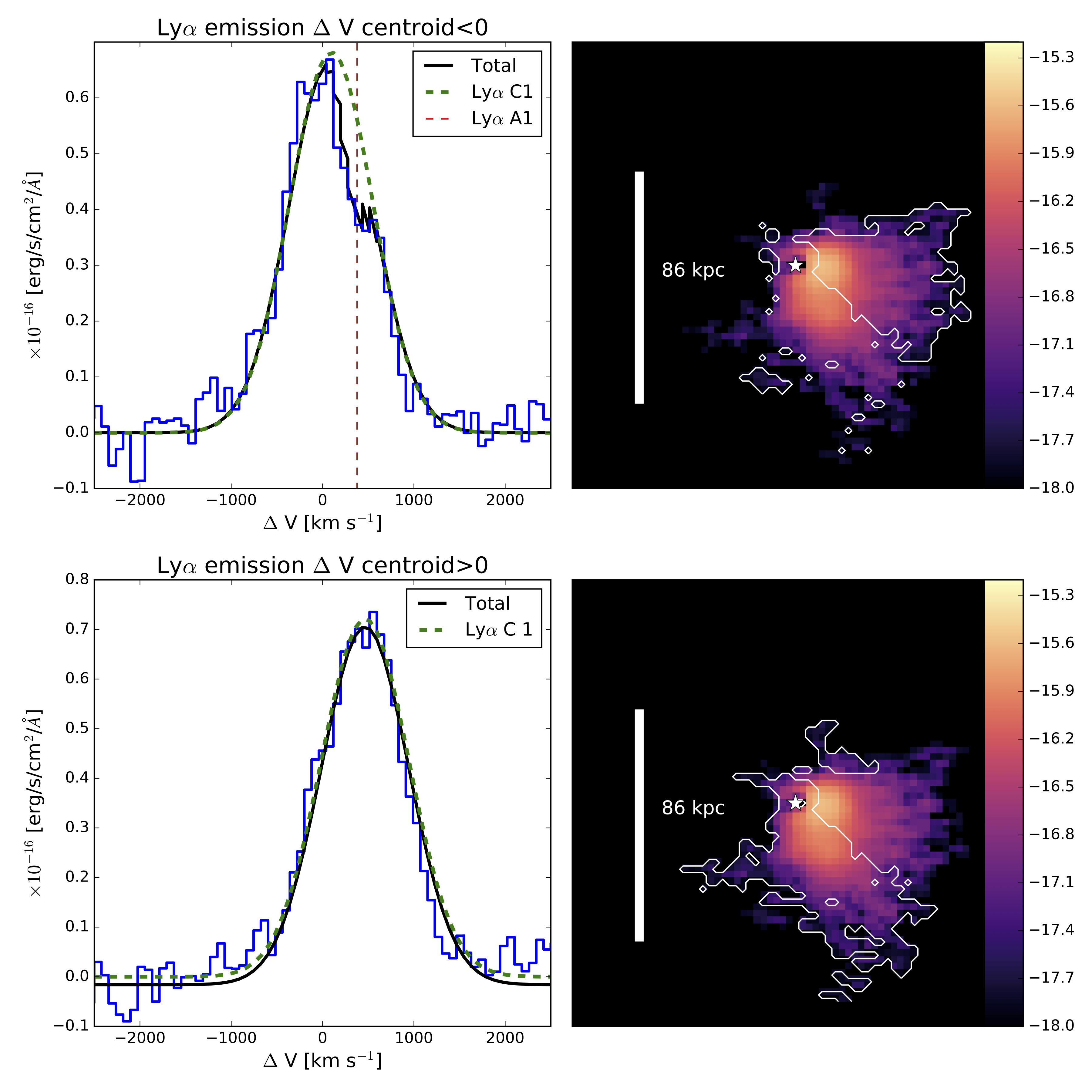}\\
    \caption{Spectra of distinct regions and kinematic components extracted from the PSF subtracted KCWI data cube of 4C 09.17. Top panel shows the \heii\ emission extracted of the \heii\ nebula and is used to derive the redshift of the CGM and identify the distinct kinematic components. Middle and bottom left rows show spectra of \ly\ extracted over a polygon region containing the lowest surface brightness contours shown on the right integrated intensity \ly\ map. Middle and bottom rows show spectra of \ly\ extracted over the two distinct blue and redshifted kinematic components over their entire, respective \ly\ emission map. Each emission line is fit with a single or multiple Gaussian components shown with a dashed line, while the total best fit consisting of the individual components is shown in black. The dashed vertical line shows absorption in the blueshifted \ly\ profile due to the foreground gas in the redshifted kinematic component. North is up and east is to the left.}
    \label{fig:spec_4C0917}
\end{figure}

To constrain the systemic velocity of the CGM surrounding the galaxy group, we integrate over the extended \heii\ emission. The spectrum shows a double-peaked emission-line profile in \heii, suggesting the presence of two kinematic components. We fit the spectrum with a sum of two Gaussians and measure a weighted flux redshift of 2.10879 and take this to be the systemic redshift of the CGM in the 4C 09.17 group. We measure the velocity relative to He II 1640 \AA\ because this is a recombination line; hence it does not suffer from resonant scattering like \ly, and can provide the true kinematics of the gas in the CGM. Following this, we also integrate over the entire individual blue/redshifted extended \ly\ emission and present them in Figure \ref{fig:spec_4C0917}. The velocities are measured relative to the frame of the CGM determined from the flux-weighted redshift described above (Figure \ref{fig:spec_4C0917}).

The high optical depth of \ly\ can allow us to probe the 3D structure of the CGM for certain geometric configurations \citep{Wang21}. For example if two or more gas filaments line up and overlap along the line of sight, then the background filament can have a portion of its emission absorbed by the foreground filament imprinting the 3D structure onto the \ly\ line profile. We detect an absorption in the \ly\ line in the blueshifted kinematic component at the velocity offset of the redshifted kinematic component as measured with the \heii\ emission line (Figure \ref{fig:spec_4C0917}, middle-left panel). This absorption indicates that the redshifted kinematic component is in the foreground since it absorbs a portion of the blueshifted kinematic component. Based on this geometric setup, the redshifted gas moves inwards towards the galaxy group. Since the light emitted from the blueshifted component is getting absorbed, this further means that the blueshifted gas is in the background. Therefore, the blueshifted component must also be moving towards the galaxy group. We discuss our fitting procedure of both the emission and absorption lines in Appendix \ref{sec:appendix_fit}. The absorption likely exists over the entire extended \ly\ halo. Even after removing spaxels near the brightest portion of the \ly\ nebula near galaxy B, we still see an absorption component. The absorber is also detected in the \civ\ line, however the spectral resolving power and SNR does not allow the same level of fitting as for the \ly\ line.

We interpret the blue and redshifted kinematic components (Figure \ref{fig:4C0917_LY_maps}) likely to be a part of two separate filaments in the CGM of the 4C 09.17 group system similar to what has been found in hydrodynamical simulations of the CGM in massive dark matter halos \citep{Stewart17} on similar spatial scales to our observations. We find the galaxy group to reside near the boundary/intersection of the two kinematic components. The blueshifted velocity of 4C 09.17 B indicates that the galaxy is likely associated with the blueshifted kinematic components of the CGM emission, while 4C 09.17 C and D are likely associated with the redshifted component. Interestingly, \heii\ and \civ\ emission encompass galaxies B and D along with the quasar host galaxy. Since the strength of recombination radiation is proportional to the electron and hydrogen density the strong detection of \heii\ could be an indication of a larger concentration of gas in the CGM near the densest portion of the galaxy group where there is a larger accumulation of gas from accretion and stripping processes. 

The morphology of the extended \ly\ emission is very interesting, showcasing a substructure consisting of narrow gas streams extending radially outwards from the galaxy group. We apply an unsharp mask with a radius of 7 pixels (2\arcsec) to the optimally extracted \ly\ emission map to highlight the gas streams, which we present as white contours in Figure \ref{fig:4C0917_system}. After applying the unsharp mask we detect the streams more clearly against the more diffuse \ly\ emission. The narrow streams are present in both kinematic components associated with the satellite galaxies 4C 09.17 B and D. For 4C 09.17 B, we also detect more diffuse emission in both the 50 mas plate scale OSIRIS observation \citep{Vayner21b} and with the larger, 100 mas plate scale observation part of this study (Figure \ref{fig:4C0917_system}, red contours). The diffuse emission seen in \oiii\ on the 20 kpc scale appears to be offset generally in the western direction. It shows stream-like structure similar to that seen in the \ly\ halo substructure. The velocity measured in \oiii\ is similar in both offset and dispersion to the extended \ly\ and \heii\ emission. The similarity of the west-ward extent of \oiii\ and \ly\ in the blueshifted kinematic component, together with similar velocity offset and elongated clumpy substructure, leads us to believe that they are part of the same warm-ionized gas structure of the CGM. Emission line ratios using the emission lines \oiii, \ha\ and \nii in \citet{Vayner21b} revealed that the more diffuse emission in 4C 09.17 B is consistent with quasar photoionization, the clumpier emission is consistent with photoionization from young stars. We interpret this as likely being due to differences in the gas column densities of the star-forming clumps vs. the more diffuse gas where ionizing photons from the quasar can more easily penetrate. 

\subsection{7C 1354+2552}

7C 1354+2552 is a luminous quasar with a bolometric luminosity of $(2.75\pm0.11)\times10^{46}$ \ergs\ at z = 2.032, measured based on the quasar narrow-line region. The host galaxy of the quasar consists of a star-forming disk galaxy with a star formation rate of 29 $\pm$3 \myr, based on the \ha\ emission-line luminosity \citep{Vayner21b}. Through modeling of the rotating galactic disk, we measure the position angle of the semi-major axis to be 75.68 $\pm$0.47 \deg East of North with a maximum line-of-sight velocity along the major kinematic axis of 309.84 $\pm$ 20.47 \kms and a velocity dispersion of 61.3 $\pm$ 7.9 \kms. The blueshifted component of the galactic disk is towards the south-eastern direction, while the redshifted is found towards the northwest.

In \citet{Vayner21b} we discovered a companion galaxy (7C1354+2552 B) towards the southeast direction. This galaxy is detected in the \oiii, \ha, and \nii emission lines. The line ratios are consistent with a star-formation as the primary source of gas photoionization with a rate of 30 \myr. In our recent ALMA band 4 observations of this system, we detected CO (4-3) emission within $\pm$ 500 \kms\ from the systemic redshift of the quasar, associated with galaxies in the KCWI field of view. 

\begin{figure}[!th]
    \centering
    \includegraphics[width=3.4 in]{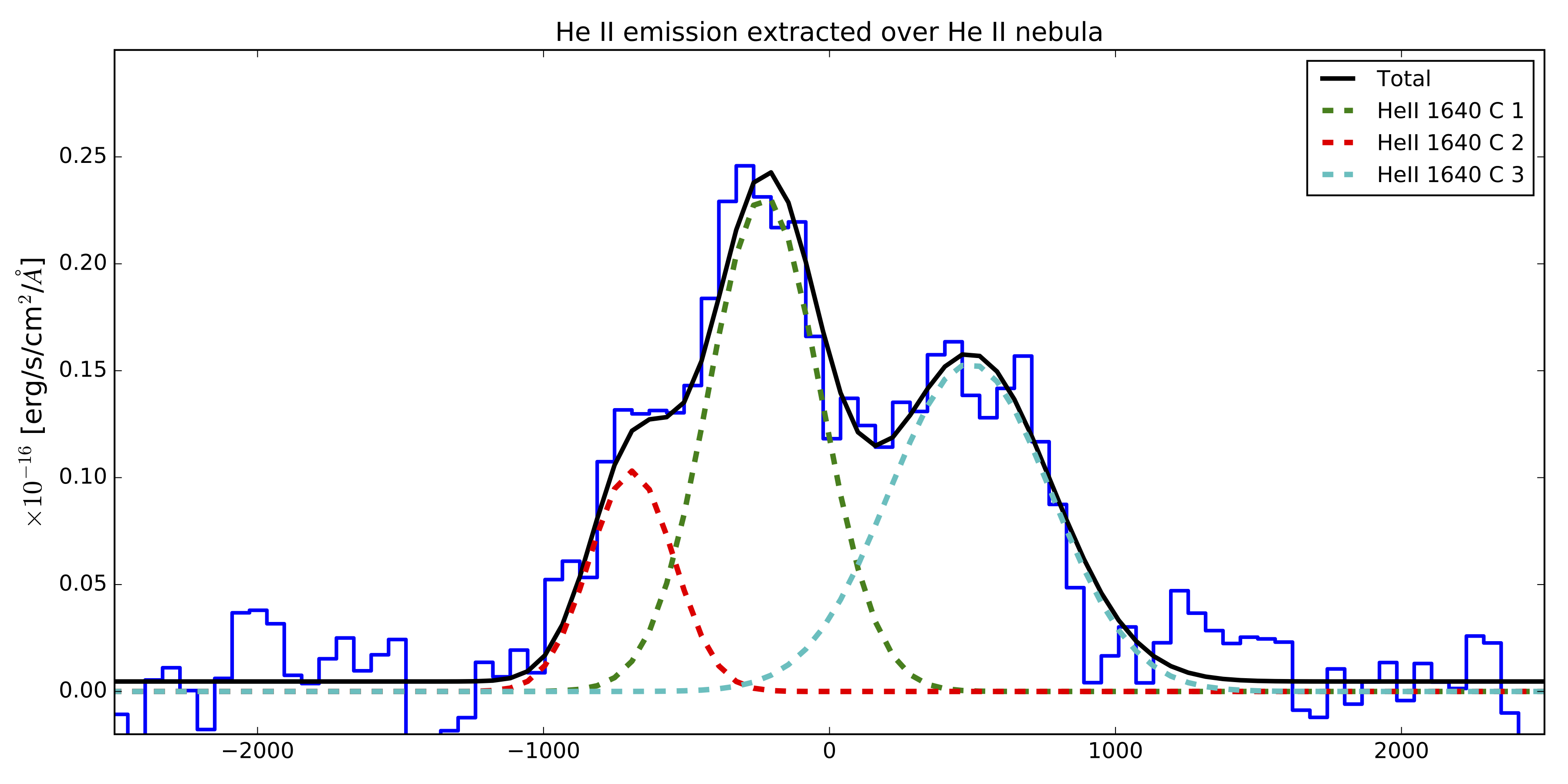}\\
    \includegraphics[width=3.5 in]{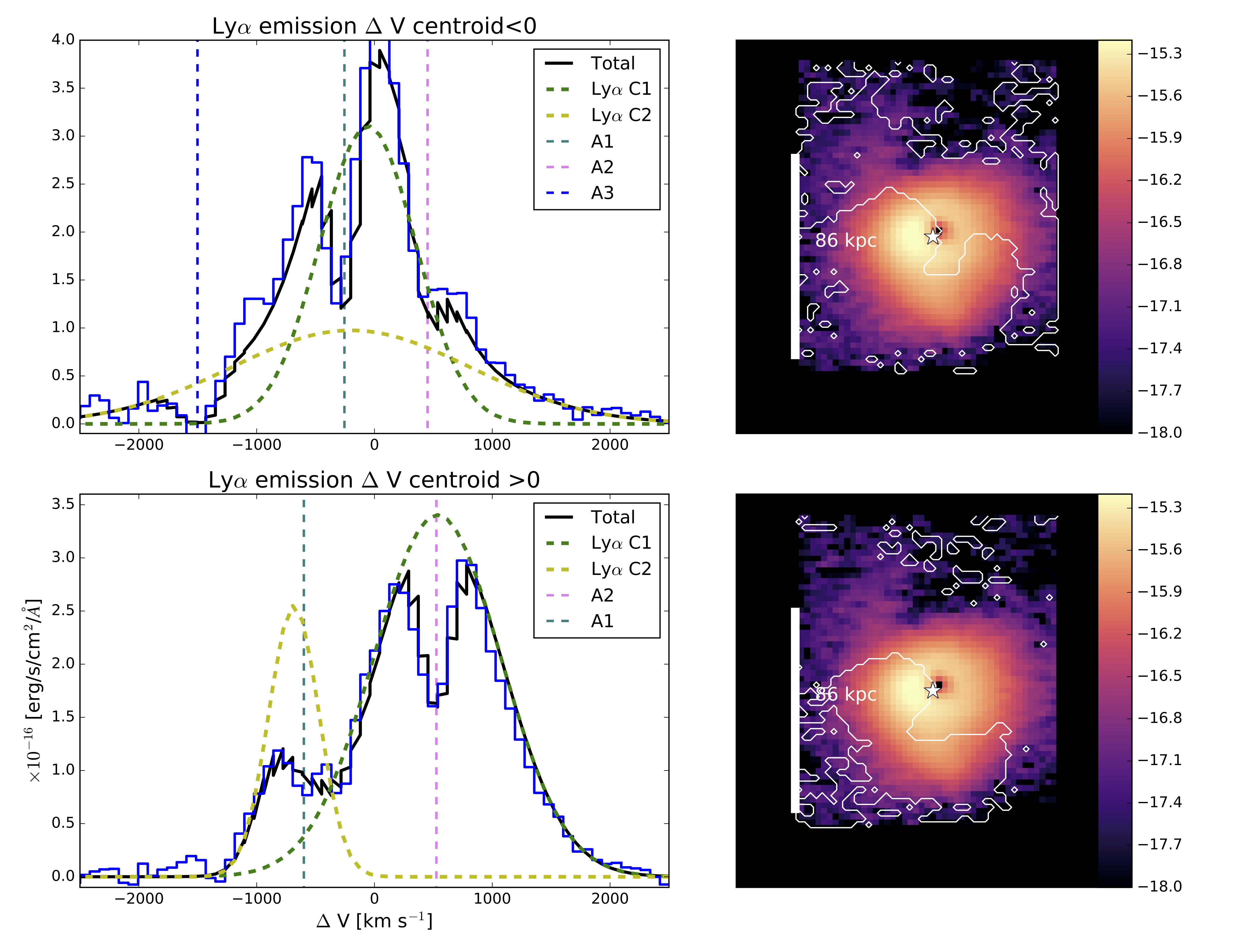}\\
    \caption{Spectra of distinct regions and kinematic components extracted from the PSF subtracted KCWI data cube of 7C 1354+2552. Top panel shows the \heii\ emission extracted of the \heii\ nebula and is used to derive the redshift of the CGM and identify the distinct kinematic components. The middle and bottom left rows show spectra of \ly\ extracted over a polygon region containing the lowest surface brightness contours of the right column map. The two rows show spectra of \ly\ extracted over the two distinct blue and redshifted kinematic components over their entire, respective \ly\ emission map. Each emission line is fit with a single or multiple Gaussian components shown with a dashed line, while the total best fit consisting of the individual components is shown in black. The dashed line in shows absorptions found in the extended \ly\ maps of both kinematic components. North is up and east is to the left.}
    \label{fig:spec_7C1354}
\end{figure}

\begin{figure}[!th]
    \centering
    \includegraphics[width=3.6 in]{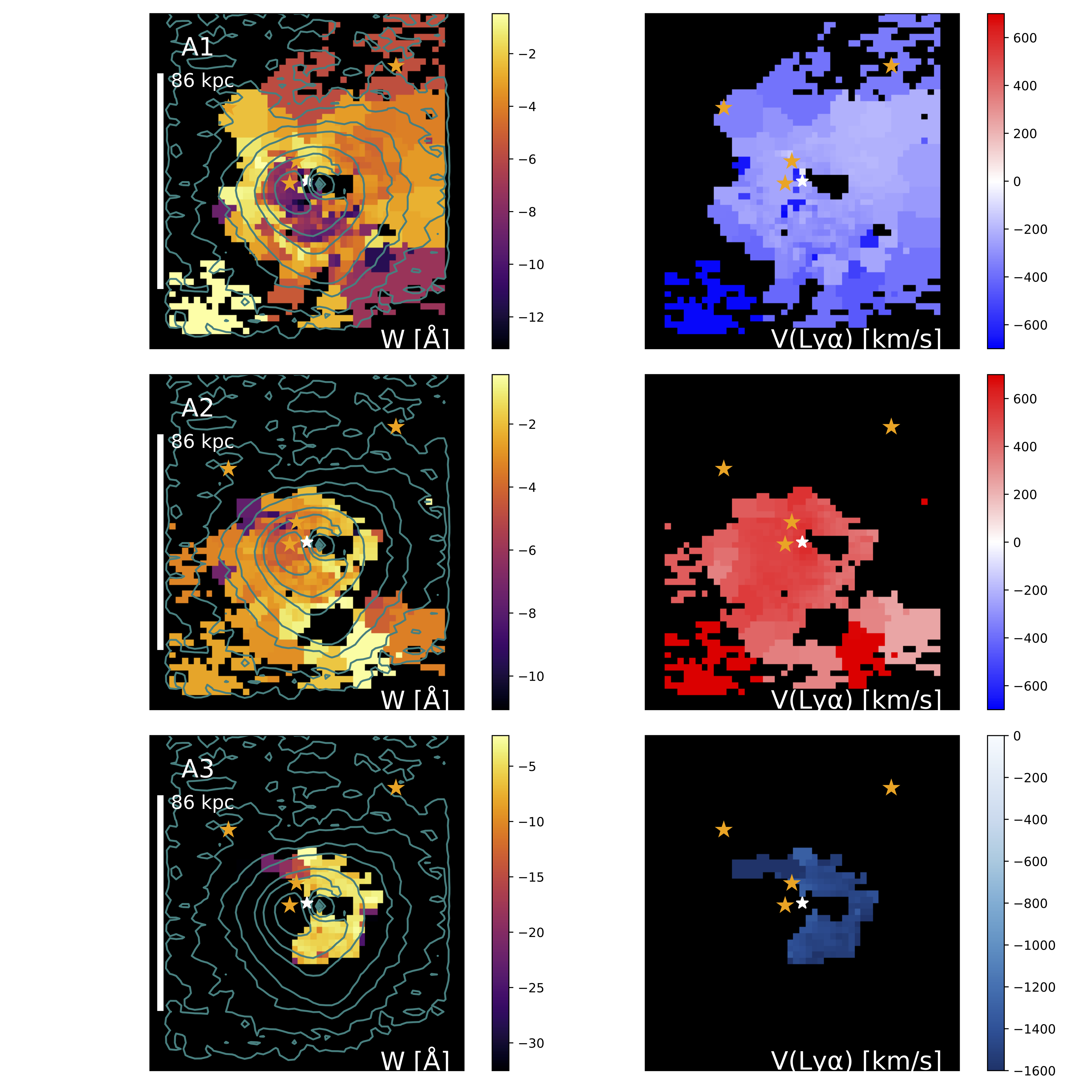}
    \caption{Equivalent width (W) and radial velocity offset maps of the \ly\ absorbers ``A1-A3" detected across the 7C 1354 +2552 \ly\ halo. Teal contours represent the \ly\ surface brightness map. Orange stars represent the location of nearby companion galaxies. North is up and east is to the left.}
    \label{fig:7C1354_abs_resolved}
\end{figure}

With KCWI, we detect extended \ly\ emission to a maximum extent of 90 kpc while we also detect \heii\ and \civ\ with a maximum extent of 50-60 kpc. The extent of the \ly\ line is likely larger and we are limited by the KCWI field-of-view. In addition we detect fainter emission lines such as Si IV 1393.755 \AA\ and OIII] 1660.809, 1666.150 \AA\ to 30 kpc extent. Surface brightness maps of the additional fainter emission lines are shown in the Appendix Figure \ref{fig:additional_lines}. Similar to 4C 09.17, we detect a gradient-like feature in the moment 1 map of all detected UV emission lines with a velocity range of -500 to 500 \kms\ and a major kinematic axis along the northwest to southeast direction. Minor differences in the velocity structure of the moment 1 map of \ly\ and \heii\ in the inner 1-2\arcsec\ likely arise due to some PSF subtraction noise or differences in the mechanisms of how the lines are produced. We extract the spectra in the kinematically distinct regions seen in the moment 1 map (Figure \ref{fig:7C1354_LY_maps}) and fit them with and fit the spectrum with a combination of a Gaussian emission line and absorption (Figure \ref{fig:spec_7C1354}). There is evidence for absorption in \ly\ that is detected over the entire extended emission region highlighted in Figure \ref{fig:spec_7C1354} associated with three different absorbers. Similar to 4C 09.17 we find the blueshifted kinematic component shows absorption in \ly\ at the velocity of the redshifted kinematic component. Similar to the case of 4C 09.17 the absorption in the spectrum of the blueshifted component in the \ly\ line at the velocity of the redshifted kinematic component indicates that we are indeed detecting inflowing gas, as both kinematic components are moving inwards towards the quasar host galaxy and galaxy group. We were able to spatially map the extent of each absorber. In Figure \ref{fig:7C1354_abs_resolved} we present a map of the resolved equivalent width and radial velocity offset of the absorbers relative to the systemic redshift of the CGM.

\begin{figure*}[!th]
    \centering
    \includegraphics[width=5.0 in]{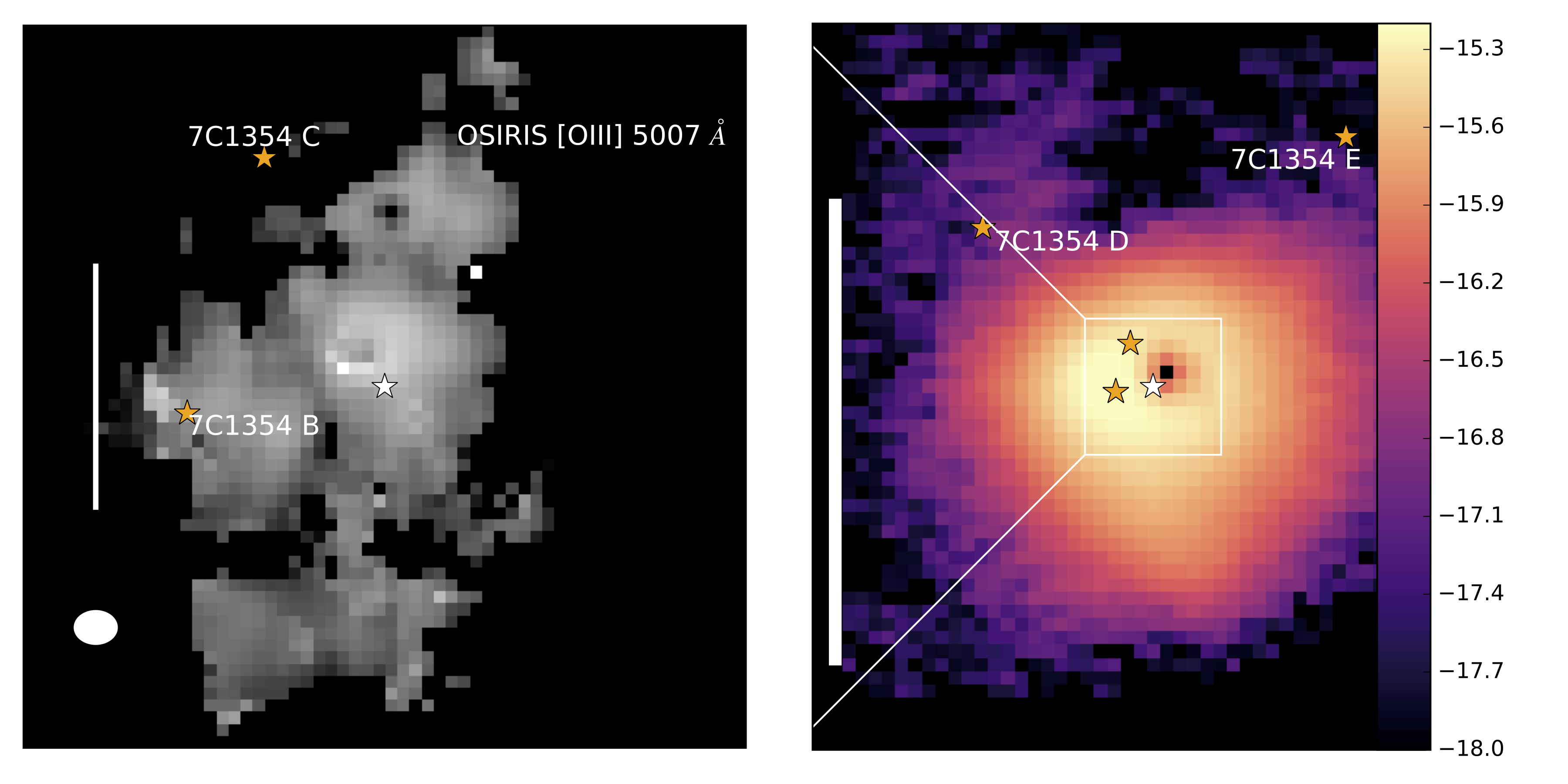}
    \caption{Detection of emission from gas components in the 7C1354+2552 system on scales ranging from 1 to 100 kpc. Left panel shows the inner most emission detected with Keck/OSIRIS IFS observations mapping the \oiii\ emission line on kpc scales. Right panel shows extended \ly\ halo associated with the group of galaxies. The orange stars represent the position of satellite galaxies detected with either ALMA or OSIRIS observations, white star represents the location of the quasar host galaxy. Box on the right represents the OSIRIS field of view on the left. North is up and east is to the left.}
    \label{fig:7C1354_system}
\end{figure*}

We present the location of the galaxies relative to the quasar host galaxy and to their position within the \ly\ halo in Figure \ref{fig:7C1354_system}. The two closest galaxies to the quasar are found to reside within the \heii\ halo. The velocity of galaxy 7C1354+2552 B indicates that it is likely associated with the redshifted kinematic component detected in \ly\ and \heii\ towards the south-west. 7C1354+2552 C and D appear to be linked by a ``bridge" structure towards the northwest, and their velocity offsets are in general agreement with the velocity of \heii\ and \ly\ found in the moment 1 map towards the northwest. Similarly a ``bridge" structure towards the north-east links the three galaxies near the centroid of the nebula with the satellite galaxy 7C1354+2552 E.

The presence of galaxies within the rest-fame UV emission-line nebulae with velocities consistent with the extended gas leads us to believe that the gas is likely associated with gas accreting onto the central galaxy group. Galaxies 7C1354 D and E are likely on their path to merge with the central galaxy group. We interpret the blue and redshifted components as filaments within the CGM along which the galaxies are moving towards the central galaxy group.

\begin{table*}[]
    \centering
    \begin{tabular}{c|c|c|c|c}
         Line component & Integrated intensity & Line center & $\Delta$ V & Velocity dispersion   \\
                                    \hline
                        &  $\times 10^{-16}$ \ergs               & \AA & \kms        & \kms                  \\
                        \hline
                        
        7C1354 \ly\ C1 $\Delta V>0$   & $5.056_{-0.008}^{+0.024}$ &$3663.964_{-0.069}^{+0.013}$&$533_{-6}^{+1}$ &$539_{-1}^{+2}$\\
        7C1354 \ly\ C2 $\Delta V>0$ & $1.488_{-0.209}^{+0.013}$ &$3649.122_{-0.037}^{+0.048}$& $-684_{-3}^{4}$ & $213_{-1}^{+15}$\\
        7C1354 \ly\ C1 $\Delta V<0$   & $3.476_{-0.007}^{+0.082}$ &$3656.784_{-0.024}^{+0.095}$&$-56_{-2}^{+8}$ &$407_{-1}^{+5}$\\
        7C1354 \ly\ C2 $\Delta V<0$ & $2.710_{-0.051}^{+0.010}$ &$3655.032_{-0.397}^{+0.074}$& $-199_{-33}^{6}$ & $1010_{-3}^{+14}$\\        
        7C1354 \heii\ C1 $\Delta V>0$ & 0.11$\pm$0.01           & 4944.88$\pm$0.46  & 578$\pm$33 & 217$\pm$26 \\
        7C1354 \heii\ C1 $\Delta V<0$ & 0.265$\pm$0.03             & 4931.17$\pm$0.39 &-255$\pm$30 & 342$\pm$25  \\        
        4C09.17 \ly\ C1 $\Delta V<0$  &  $0.892_{-0.010}^{+0.016}$  & $3780.728_{-0.249}^{+0.072}$ &$116_{-20}^{+6}$ & $463_{-5}^{+17}$ \\
        4C09.17 \ly\ C1 $\Delta V>0$  &  0.990 $\pm$ 0.01           & 3785.38 $\pm$ 0.17 &465$\pm$22 & 478$\pm$14 \\
        4C09.17 \heii\ C1 $\Delta V>0$ & 0.062$\pm$0.006           & 5106.30 $\pm$ 0.9  &393$\pm$53 & 307$\pm$46 \\
        4C09.17 \heii\ C1 $\Delta V<0$ & 0.05$\pm$0.01             & 5093.98 $\pm$ 0.65 & $-329\pm42$ & 231$\pm$31 \\
                        \hline
    \end{tabular}
    \caption{Best fit emission line properties from spectra integrated over individual kinematic components}
    \label{tab:emission_table}
\end{table*}

\begin{table*}[]
    \centering
    \begin{tabular}{c|c|c|c|c}
         Line component & Column density & line center &$\Delta$V             & Doppler parameter [b]  \\
                                            \hline
                        &  log10(cm$^2$)        & \AA &\kms             & \kms                  \\
                                            \hline
        4C 09.17 \ly\ $\Delta V < 0$ A1  &  $13.482_{-0.055}^{+0.063}$ & $3783.907_{-0.151}^{+0.321}$ &$368_{-12}^{+25}$ & $166_{-19}^{+61}$  \\
        7C 1354+2552  \ly\ $\Delta V > 0$ A1  &  $14.238_{-0.085}^{+0.006}$         & $3650.189_{-0.047}^{+0.466}$ &$-597_{-4}^{+38}$ & $349_{-6}^{+2}$  \\
        7C 1354+2552  \ly\ $\Delta V > 0$ A2  &  $13.795_{-0.003}^{+0.004}$         & $3663.871_{-0.014}^{+0.007}$ & $525_{-1}^{+1}$& $123_{-2}^{+1}$  \\
        7C 1354+2552  \ly\ $\Delta V < 0$ A1  &  $13.983_{-0.003}^{+0.006}$         & $3654.352_{-0.007}^{+0.010}$ &$-255_{-1}^{+1}$ & $97_{-3}^{+1}$  \\
        7C 1354+2552  \ly\ $\Delta V < 0$ A2  &  $13.735_{-0.004}^{+0.010}$         & $3662.954_{-0.026}^{+0.020}$ & $450_{-2}^{+2}$& $109_{-1}^{+12}$  \\
        7C 1354+2552  \ly\ $\Delta V < 0$ A3  &  $15.821_{-0.101}^{+0.053}$         & $3639.144_{-0.331}^{+0.052}$ & $-1503_{-27}^{+4}$& $83_{-1}^{+15}$  \\        
                        \hline
    \end{tabular}
    \caption{Best fit absorption line properties from spectra integrated over individual kinematic components}
    \label{tab:absorption_table}
\end{table*}
\pagebreak
\section{Discussion}
\label{sec:discussion}
\subsection{Evidence for gravitational motion and gas accretions in the CGM}
Both sources in this study show a well-defined gradient structure in the radial velocity maps of \ly, \heii, and \civ. The fact that a similar velocity structure is observed in the \heii\ recombination line leads us to believe that these kinematic structures are not caused by radiative transfer effects of \ly\ and are indeed real gas motion in the CGM of these two systems. In both objects the blueshifted and redshifted kinematic components are associated with nearby satellite galaxies. The gradient patterns do not appear to be associated with large scale structure of the quasar host galaxies in either of the systems. In fact for 7C 1354+2552 the gradient pattern seen in the moment 1 map is counter to the rotational pattern seen in the quasar host galaxy. We further discuss the significance of of the angular momentum misalignment later in the discussion section. \\

For both systems, we have defined a systemic redshift for the gas in the CGM based on the luminosity weighted centroid of the blue and redshifted kinematic components measured over the entire \heii\ halo. In both systems, we detect a concentration of 2-3 galaxies of similar mass within a 20 kpc radius from the quasar, indicating that the quasar host galaxies of 4C 09.17 and 7C1354+2552 reside in a group environment. The \heii\ emission appears to be concentrated in the galaxy group, potentially near the node of the dark matter halo. Extended \heii\ emission has recently been found around other high redshift quasars, and these systems also show similar results where \heii\ is detected near a concentration of galaxies \citep{Cai17,Cantalupo19,Herenz20,Husemann21}. In both 4C 09.17 and 7C1354+2552, the blueshifted and redshifted kinematic components appear to move towards the central galaxy group. For both sources, the presence of absorption in the \ly\ line gives us clues to the three-dimensional structure of the gas in the CGM, where the redshifted component is located in front of the blueshifted kinematic component, meaning that both kinematic components are moving towards the central concentration of galaxies. While \ly\ emission line can suffer from several radiative transfer effects that cause arbitrary broadening and velocity shifts, the optical depth helps with interpretation of relative gas motion in the CGM in high signal to noise ratio data where we can both resolve the gas in the CGM and measure the emission and absorption profile. While we detect signs of inflows based on \ly\ absorption kinematics, similar analysis of a radio-loud galaxy found evidence for outflowing material on CGM scales based on resolved \ly\ absorption \citep{Wang21}.\\

Based on statistical analysis, we know that quasars reside in dark matter halos of $10^{12-13.7}$\msun \citep{White12,hall18a} at z$\sim2$. For an NFW dark-matter profile \citep{Navarro96}, such halos are expected to have virial velocities of 200-400 \kms \citep{White12,Buckley14,Shull14}, the observed motion in both of the kinematic components are consistent with gravitational motion in a massive dark matter halo. The observed gradients are similar to those found in other high redshift CGM around luminous quasars at $z=2-3$. Other observational works have associated these kinematic structures as inflowing material from the CGM \citep{Arrigoni-Battaia18,mart19,Arrigoni-Battaia21} based on the similarity between velocity offsets and similar gradient-like structures found in hydrodynamical simulations of massive dark matter halos \citep{Stewart17}. 

\subsection{Deriving the warm-ionized gas mass using \ly\ and inflow rates}
It is interesting to ask at what rate the cold gas in the CGM is accreting onto the galaxy groups in both systems. However, first, we need to measure the amount of warm-ionized gas in the CGM. Over the last ten years, several works have estimated the amount of warm ionized gas in the CGM around quasars using \ly\ as a tracer of the gas mass. The first method assumes that \ly\ emission arises in regions that are optically thin to Lyman continuum photons. Using a spectral model where the quasar is the primary source of ionization with the assumption that the majority of the gas is in the ionized state, \citet{Hennawi13} find the following relationship between the hydrogen column density and the average surface brightness of \ly:

\begin{equation}
    \frac{N_{H}}{10^{20}\rm cm^{-2}} = \frac{SB}{7.7\times10^{-19}}\left(\frac{1+z}{3.0}\right)^{4}\left(\frac{f_{c}}{1.0}\right)^{-1}\left(\frac{n_{H}}{0.1 {\rm cm}^{-3}}\right)^{-1}
    \label{eq:optically_thin_ly}
\end{equation}

\noindent where $f_{c}$ is the covering factor for the clouds in the CGM and $n_{H}$ is the hydrogen number density, which equals the electron density $n_{e}$ under the assumption that the majority of the gas is ionized. The above equation holds true for neutral hydrogen column densities of $N_{HI}\ll10^{17}$cm$^{-2}$. The electron or hydrogen density is always a major uncertainty when calculating ionized gas masses \citep{Harrison18}. Based on statistical observations from the line of sight absorption measurement in the CGM of high redshift luminous quasars, the median total hydrogen column density is $log_{10}(N_{H})\sim20.5\pm1$ within a projected radius of 100-200 kpc \citep{Lau16} and the average covering factor is 0.5 \citep{Prochaska13a}. The average \ly\ surface brightness within 100 kpc, corrected for redshift surface brightness dimming is 1$\times10^{-17}$\ergs arcsec$^{-2}$, based on statistical observations of the CGM around luminous quasars \citep{Cai19}. Plugging these two values into equation \ref{eq:optically_thin_ly} we can estimate that the expected average hydrogen number density is 1 cm$^{-3}$. Multiplying equation \ref{eq:optically_thin_ly} by the surface area of the \ly\ emitting region can provide us with a hydrogen gas mass. The electron density is likely a lower limit since some annuli used to compute the average surface brightness maps do not have full emission across the entire annulus \citep{Heckman91b,Arrigoni-Battaia15b,Hennawi15}.

Another method is to assume that the CGM consists of individual clouds in the CGM that are all at the same density. The \ly\ line is also assumed to be optically thin in the case A recombination regime \citep{OsterbrocknFerland06}. 

\begin{equation}
     M_{H II}=(n_{p}m_{p} + n_{He}m_{He})Vf . 
     \label{eq:total_ionized_mass}
\end{equation}

\noindent $V$ is the volume of the gas emitting region in the CGM, $f$ is the volume filling factor (the ratio of the volume of emitting clumps to the total volume of the region), and $n_{p}$ is the proton number density. $n_{He}$ and $m_{He}$ are the number density of helium and the mass of a helium atom. We assume a solar abundance for helium in the CGM gas. We further assume the gas to be fully ionized where helium is an equal mix of HeII and HeIII. Under these assumptions, we get the following relationships:

\begin{equation}
\begin{aligned}
    n_{He} &= 0.1 n_{p}\\
    n_{e} &= n_{p} + \frac{3}{2}n_{He}\\
    n_{e} &= 1.15 n_{p}
\end{aligned}
\label{eq:relationships}
\end{equation}
\noindent
Following from \citet{OsterbrocknFerland06}, we can write the line luminosity due to recombination:
\begin{equation}
    L(Ly\alpha) = n_{e}n_{p}j_{Ly\alpha}Vf
    \label{eq:ly_recombination}
\end{equation}

\noindent where $n_{e}$ is the electron density, $n_{p}$ is the proton density and $j_{Ly\alpha}$ is the emissivity. We use an emissivity value of 1.53$\times10^{-24}$ erg~cm$^{3}$~s$^{-1}$, calculated using an electron density of 1 \eden\ and gas temperature of 20,000 K using the \textit{PyNeb} package \citep{Luridiana15}. By combining equation \ref{eq:total_ionized_mass}, \ref{eq:relationships} and \ref{eq:ly_recombination} we can derive the following equation for the relation between, \ly\ luminosity, electron density and hydrogen mass:

\begin{equation}
    M_{H II} = 7.7\times10^{9}M_{\odot}\frac{L_{Ly\alpha}}{1\times10^{43} \mbox{ erg  s}^{-1}}\left(\frac{n_e}{1\mbox{ cm}^{-3}}\right)^{-1}.
    \label{eq:ionized_gas_mass_ly}
\end{equation}

Both \ly\ mass derivation methods provide similar results. In Table \ref{tab:hydrogen_mass} we present the amount of warm-ionized gas mass derived using equation \ref{eq:ionized_gas_mass_ly} for the total \ly\ nebula and only for the portion where we detect \heii.

\subsection{Deriving the warm-ionized gas mass using HeII 1640 and inflow rates}
For \ly\, the combination of resonant scattering, photoionization due to quasar and star formation, and collisional excitation due to gravitational cooling makes it difficult to estimate the amount of warm-ionized gas in the CGM. There is evidence within our two nebulae that the \ly\ emission may have a strong contribution from resonant scattering due to relatively large equivalent widths observed in \ly\ absorption on the spatial extent of the \ly\ halo \citep{Hennawi13}. Furthermore, the difference in the velocity dispersion between \ly\ and \heii\ after correcting for beam-smearing further indicates that resonant scattering plays a role in the \ly\ line, since the \ly\ photons scattered by the higher velocity gas can escape the nebulae more easily. It is not easy to decipher the amount of line luminosity caused by each ionization source.

Because \heii\ is optically thin, we can assume that most \heii\ emission comes from recombination and that the quasar is the primary source of gas ionization, with a minor contribution from collisional excitation. We assume each cloud has the same density and the density is constant across the \heii\ nebula, similar to our assumption for the \ly\ derived mass in equation \ref{eq:ionized_gas_mass_ly}.

Using the formulation of \cite{OsterbrocknFerland06}, the \heii\ luminosity due to recombination is given by:

\begin{equation}
    L(He II) = n_{e}n_{He III}j_{He II 1640}Vf
    \label{eq:heii_recombination}
\end{equation}

\noindent where $n_{e}$ is the electron density, $n_{He III}$ is number density of doubly ionized helium. $j_{He II 1640}$ is the emissivity of the 1640 \AA\ \heii\ emission line, under the assumption of case B recombination at the lower density. We use an emissivity value of 5.36$\times10^{-24}$ erg~cm$^{3}$~s$^{-1}$, calculated using an electron density of 1 \eden\ and gas temperature of 20,000 K using the \textit{PyNeb} package \citep{Luridiana15}. Combining equations (\ref{eq:total_ionized_mass},\ref{eq:relationships},\ref{eq:heii_recombination}) we obtain the following total H II ionized gas mass - $L_{He II 1640}$ relationship:

\begin{equation}
    M_{H II} = 4.4\times10^{10} M_{\odot} \frac{L_{He II}}{1\times10^{43} \rm~erg s^{-1}}\left(\frac{n_e}{1\mbox{ cm}^{-3}}\right)^{-1}
    \label{eq:ionized_gas_mass_heii}
\end{equation}

In table \ref{tab:hydrogen_mass} we derive the ionized gas mass within the maximum extent of \heii\ using only spaxels where \heii\ is detected. In all cases, we assume an electron density of 1 cm$^{-3}$.

\begin{table*}[]
    \centering
    \begin{tabular}{c|c|c|c|c|c|c}
        
         Object & SB \ly\ [R max] & M$\rm_{H II}$ \ly  [R max] & SB \ly\ [R \heii] & M$\rm_{H II}$ \ly [R \heii] & SB \heii [R \heii] & M$\rm_{H}$ \heii [R \heii]\\ 
         \hline
         7C1354 & 6$\pm0.7$ & 27$\pm$3 & 18 $\pm$ 2 & 23 $\pm$2 & 0.8$\pm$0.1 & 6$\pm$0.5  \\
         4C 09.17 & 2.4$\pm0.2$ & 5$\pm$0.5 & 9.3 $\pm$ 0.9 & 3 $\pm$0.3 & 0.5$\pm$0.1 & 1.0$\pm$0.1

    \end{tabular}
    \caption{SB \ly\ [R max] is the average surface brightness over the entire nebula, out to the maximum detected extent in units of 1$\times10^{-17}$ \ergs arcsec$^{-2}$. SB \ly\ [R \heii] is the average surface brightness over the spaxels where \heii\ is detected in units of 1$\times10^{-17}$ \ergs arcsec$^{-2}$. SB \ly\ [R \heii] is the \heii\ surface brightness in units of 1$\times10^{-17}$ \ergs arcsec$^{-2}$. M$\rm_{H II}$ is the total H II mass derived from equation \ref{eq:ionized_gas_mass_ly} and \ref{eq:ionized_gas_mass_heii} for \ly\ and \heii, respectively, in units of 1$\times10^{10}$\msun.}
    \label{tab:hydrogen_mass}
\end{table*}

To estimate the ionized gas inflow rate, we divide the gas mass by the dynamical time scale ($R/v_{r}$) of the inflowing material. For the radius ($R$), we use the maximum extent of the \heii\ surface brightness profile measured down to 2$\sigma$, and for the velocity ($v_{r}$) we use the luminosity weighted velocity difference between the red and the blueshifted kinematic components. We estimate inflow velocities of 181 \kms\ and 180 \kms\ for 4C 09.17 and 7C1354+2552, respectively. These inflow velocities are consistent with motion expected in a massive dark matter halo \citep{White12} and similar to the expected inflow velocity based on cosmological cold-gas inflows found in \citet{Goerdt15,Beckmann17}. For 4C 09.17, we obtain an estimated inflow rate of 60 \myr, while for 7C 1354+2552, we obtain a value of 200 \myr. These inflow rates are also consistent with the expected value from hydrodynamical simulations \citep{Goerdt15,Beckmann17}. These values are an order of magnitude estimate. Several factors are unknown, such as the geometry of the inflowing matter, the electron density of the gas in the CGM, the assumption of constant electron density across the nebulae, unknown power mechanism of \ly\ emission, unknown temperature of the gas producing \ly\ and \heii\ emission, and the unknown fraction of HeIII. Measuring the electron density of the gas in the CGM is challenging with current instruments, especially at the lower electron density of the gas in the CGM. 

We notice a significant difference between the ionized gas mass derived from \ly\ and \heii, over the same aperture. In both cases, the mass derived from \ly\ is a factor of 3-4 greater than what we estimate in \heii. A likely scenario, as discussed earlier, is that a considerable fraction of the \ly\ emission comes from resonant scattering. As noted in the detailed photoionization simulation in \citet{Hennawi13} the surface brightness in \ly\ due to resonant scattering can be very similar to recombination radiation from either optically thin or thick gas conditions. A significantly smaller gas reservoir can produce scattering emission with the same surface brightness as recombination from a much larger gas mass. Using the known equivalent widths that we measure in the extended \ly\ halo for the three detected absorbers in 7C1354+2552 we can roughly estimate the expected surface brightness value in the \ly\ line from scattering. Using equation 20 from \citet{Hennawi13}, for equivalent width of 6-8 \AA\ within 40 kpc from the quasar we expect a surface brightness in \ly\ due to scattering on the order of 3-4$\times10^{-17}$\ferg $\rm arcsec^{-2}$, which is close to the observed surface brightness within 40-80 kpc. Hence a large fraction of the \ly\ emission can be due to resonant scattering and if its not properly taken into account can drastically overestimate the amount of warm ionized gas in the CGM.

The \heii\ line comes almost entirely from recombination and hence likely traces the larger gas reservoir. This showcases the importance of using a recombination line when measuring the warm ionized gas mass in the resolved CGM. Observations through hydrogen Balmer lines can also help, in addition to the \heii\ data. Likely our assumption of a single constant electron density for the gas in the CGM does not hold. It may be partially responsible for the differences in the gas masses derived from \heii\ and \ly; however, it is difficult to quantify this uncertainty with our current data set. Most likely we are still seeing small dense clouds within presumably lower density, hotter, volume filling gas.

Nevertheless, the gas masses derived from both lines are likely lower limits on the total gas mass in the CGM as we are still missing tracers of the neutral atomic, molecular \citep{Ginolfi17}, and extremely hot ionized medium gas \citep{Gobat19}. The estimated inflow rates are for each entire galaxy group, assuming they reside in one coalesced dark matter halo. The inflow rates on the individual galaxies are likely lower; however, if the galaxies merge with the central quasar host galaxy, then the current inflow rate can be thought of as the total baryon matter supply for the central cluster/group galaxy.

\subsection{Evidence for gas stripping in 4C 09.17 system.}

Galaxy interactions in proto-groups and clusters create a hot virialized halo with temperatures up to 10$^{7}$ K. In addition, feedback from supermassive black-holes through energy-conserving shocks can help provide the halo with hot diffuse gas that does not radiate energy efficiently \citep{Faucher12,Zubovas12,Zubovas14}. Since it does not radiate efficiently, detecting the hot gas associated with quasar-driven winds is difficult. However, it can be indirectly constrained by studying the dynamics of the outflows. One consequence of energy-conserving shocks is that they produce a multi-phase outflow with a momentum flux ratio between momentum flux of the outflow and radiation momentum flux of the accretion disk $>$ 2 on kpc scales. In the case of 4C 09.17, we have recently detected an outflow that is likely driven by an energy-conserving shock \citep{Vayner21d}, indicating the presence of a hot gas medium. Ram pressure due to the hot gas can cause stripping and is a common byproduct in local clusters and can be seen as a large extended gas streams extending from satellite galaxies. These galaxies are often referred to as ``Jellyfish" galaxies \citep{Ebeling14}. The morphology of 4C 09.17 B combined with the extended streams structure likely indicates that we have detected stripping of gas from the galaxy as it merges with the quasar host galaxy. The gas in 4C 09.17 B along with the extended gas streams are similar morphologically to a galaxy in a recent FOGGIE simulation (Cyclone halo) that is being stripped through ram pressure in a dark matter halo with a mass of 10$^{12}$\msun \citep{Simons20}. The redshifted kinematic component in 4C 09.17 associated with galaxy D shows similar streams morphology, likely both galaxies are experiencing some level of gas stripping as they move down along with the accretion flow.

As discussed in \citet{Angles-Alcazar17} stripping of gas from satellite galaxies can provide a large reservoir of material that can re-accrete onto the central galaxy or the group/cluster node. The stripped material from the 4C 09.17 B galaxy is unlikely to escape the potential of the galaxy group and will likely re-accrete onto the galaxy group along with the rest of the CGM gas that is part of the blue and redshifted kinematic components. Multiple processes can provide the gas into the CGM of massive dark matter halos \citep{Angles-Alcazar17}. Freshly accreted and recycled gas play important roles in supplying gas into massive galaxies at high redshift. In 4C 09.17, merger activity, gas accretion and stripping occurs at similar times. Both accretion of cold gas from the CGM and exchange and transfer of gas from the ISM of the satellite galaxies are important roles in gas accretion onto the quasar host galaxy and the central galaxy groups. Based on our unsharp-mask analysis, we identify that about 24\% of the \ly\ flux is associated with the elongated stream-like structure in the blueshifted kinematic component. Under the assumption that the gas condition and electron densities are similar between the diffuse and elongated \ly\ emission, we can directly translate to a percentage of the warm-ionized gas mass that is stripped and is re-accreting onto the galaxy group vs. direct gas accretion from the CGM, indicating that a substantial fraction of the accreting gas can come from gas-stripping of material from satellite galaxies. Recent observations by \citet{Chen21} also have unidentified evidence for gas stripping in a dusty star-forming galaxy in the well-studied \ly\ halo around UM287.

\subsection{Dynamics of inflows and outflows.}

Hydrodynamical simulations predict the axis along which outflows and inflows occur. The general picture is that the two are close to 90 degrees apart \citep{tuml17}. Typically outflows occur at high galactic latitudes, near the poles, relative to the galaxy disk, while the inflows occur perpendicular, close to the plane of the galaxy. Often this scenario is referred to as the ``galactic fountain" model \citep{tuml17}. However, most of the existing evidence in support of this picture is statistical in nature and is based on absorption-line observations of galaxies and their halos against multiple background quasars. Isolating outflowing and inflowing gas in the same system is complex. It requires observations that span large spatial scales that can probe both high spatially resolved observations on a galactic scale and reach the necessary sensitivity to diffuse emission on CGM scales.

\begin{figure}
    \centering
    \includegraphics[width=3.1in]{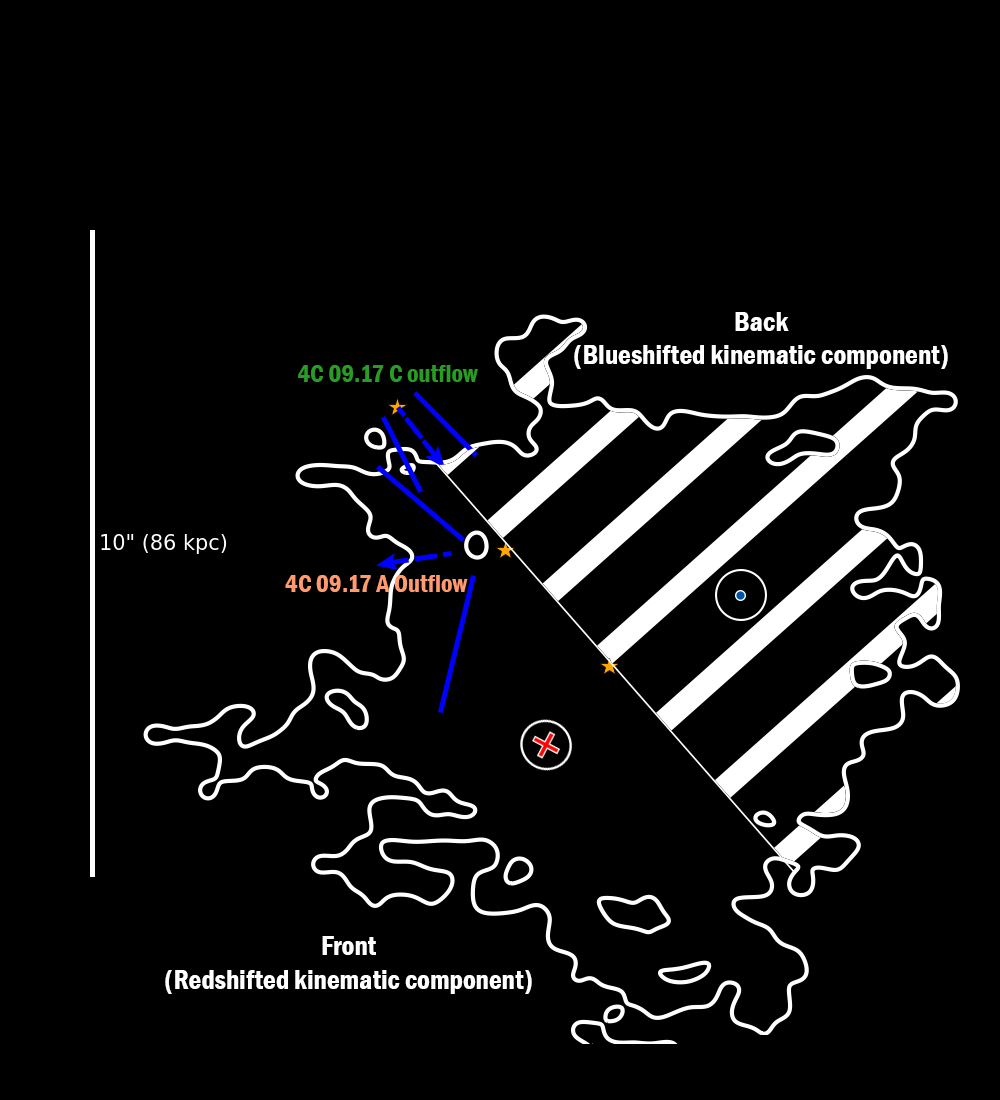}
    \caption{Schematic diagram showcasing the motion of gas in the 4C 09.17 system. Accretion from the redshifted component is shown as an in-circled ``x" moving into the image while the blue shifted component is shown as an in-circled dot moving out of the image. Dashed line showcases that the blueshifted component is located behind the redshifted component. The minor kinematic axis of the \ly\ radial velocity map bisects the two kinematic components with a white line moving through the three galaxies that are part of the quasar host galaxy group. Outflows from 4C 09.17 A and C are shown with blue lines and dashed arrows. North is up and east is to the left.}
    \label{fig:feeding_feedback_4C0917}
\end{figure}

In the case of 4C 09.17, we have evidence for both outflowing and inflowing gas. While we do not have a full three-dimensional picture of the inflowing and outflowing gas vectors, we can see if the projected directions appear to match the ``galactic fountain" model. We discovered a multi-phase molecular and ionized outflow extending towards the west, blueshifted relative to the quasar. The outflow appears to be moving away from the redshifted kinematic component of the CGM gas. The extent of the outflow is indeed close to 90 degrees apart from the major kinematic axis of the CGM gas. Interestingly, the extent and direction of the molecular outflow in 4C 09.17 C is close to the minor axis of the radial velocity map. This evidence hints at a picture where the outflows and inflows occur along different axes in the 4C 09.17 system. Based on our order of magnitude estimates, the total outflow rate in 4C 09.17 A (the quasar host galaxy) is 450 \myr, only accounting for the molecular and ionized gas phase. The inflow rate of ionized gas within 30 kpc from the quasar is 60 \myr, based on the \heii\ mass and an inflow velocity of 361 \kms.
Interestingly both numbers are comparable; however, when including the molecular outflow associated with 4C 09.17 C of 2500 \myr, then the outflow dominates, and there may be a net loss of gas from the galaxy group at the present time. For 7C 1354+2552, the inflow rate measured within 50 kpc based on the \heii\ mass is 200 \myr. We present a schematic diagram of the complex inflowing and outflowing structures in the 4C 09.17 system in Figure \ref{fig:feeding_feedback_4C0917}. In 7C 1354+2552, we have only detected outflowing gas in the ionized gas phase and near the nuclear region of the host galaxy, within 1 kpc with a rate of 52 \myr. The inflow and outflow rates in 7C1354+2552 are comparable, and there is likely more of a balance between the accretion and outflow, at least for the ionized gas phase. Measurement of the neutral gas and molecular gas \citep{Ginolfi17,Emonts18,VidalGarcia21} in the CGM and detection of outflows in more gas phases are critical in understanding the total inflow and outflow rates and the overall balance between feeding and feedback in these group systems.

\subsection{Angular momentum axis misalignment}

In 7C 1354+2552 the major kinematic axis of the \ly\ and \heii\ gradient pattern are counter to the major kinematic axis of the galaxy disc \citep{Vayner21b}, hence we do not think that is associated with a larger scale galaxy disc. Misalngment between the angular momentum distribution of the disk and the accretion flows is common in massive dark matter halos on large $>40$ kpc scales in hydrodynamical simulation \citep{Hafen22}. 

Typically the momentum distribution becomes aligned on smaller $<20$ kpc scale and is a necessary process to help create thin star forming disks in massive galaxies \citep{Hafen22}. Likely the location of where the angular momentum aligns may be unresolved in our observations, since our angular resolution is 10 kpc. However, the boundary of 0 km/s in 7C 1354+2552 radial velocity map hints at an in-spiral pattern, consistent with the expected angular momentum vector alignment predicted in hydro simulations on these angular scales. Higher angular resolution observation near the intersection of multiple kinematic components may help us observe the turn-over in the kinematic pattern as the gas accretes and falls onto galaxy discs in the early Universe providing fuel for future star formation. Interestingly, the sizes of the \heii\ emitting region roughly matches the location of where cooling flows from the CGM change angles, cool down to 10$^{4.5}$K and ``in-spiral" as the gas accretes and accumulates onto the galactic disk in massive (10$^{12}$ \msun) dark matter halos in hydrodymanical simulations \citep{Hafen22}. We speculate that the \heii\ emission may be associated with denser regions near the node where multiple CGM filaments intersect and cause an increase in the gas density, allowing for \heii\ to be more easily detected.

\pagebreak
\section{Conclusions}
\label{sec:conc}
We conducted KCWI observation of two radio-loud quasars, 4C 09.17 (z=2.1083) and 7C 1354+2552 (z=2.032), targeting the UV emission lines \ly, \civ\ and \heii\ redshifted into the optical bands to resolve and map the CGM. We found the following results:

\begin{enumerate}
    \item We detect extended \ly\ emission with a maximum extent of $\sim$ 90 kpc around both quasars.
    \item We detect extended \heii\ and \civ\ emission with maximum extents of 30-50 kpc.
    \item In 7C 1354+2552 we additionally detect extended emission in the Si IV and OIII] UV emission lines.
    \item The radial velocity maps of the UV emission lines show a gradient feature with velocity ranges of -500 to +500 \kms. Combining the kinematics of the emission lines together with spatially resolved \ly\ absorption, we find that the kinematic maps consistent with inflowing gas.
    \item By combining the data with multi-wavelength observations from Keck OSIRIS and ALMA, we find that the extended \ly\ emission is associated with a group of galaxies. The \heii\ nebulae in both sources is associated with the over-density of galaxies.
    \item In the 7C 1354+2552 system, we find that the extended \ly\ emission connects a bridge between the quasar host galaxy and three galaxies detected with ALMA. This likely indicates that the gas associated with the two kinematic components in this system is also associated with filamentary gas accretion from the CGM.
    \item We use the \heii\ to estimate the amount of warm-ionized gas in the CGM within the \heii\ halo. We measure 1-6 $\times10^{10}$ \msun within 30-50 kpc from the quasar. Using the gas's dynamical time scale, we estimate an inflow rate of 60-200 \myr, within an order of magnitude of the multi-gas phase outflow rates detected in both quasar host galaxies.
    \item We find that the inflow and outflow direction are close to 90\deg\ apart in 4C 09.17, consistent with the hydrodynamical models of the gas kinematics and dynamics in the CGM of massive galaxies.
    \item We detect narrow gas streams associated with companion galaxies in the 4C 09.17 system that point radially outwards from the quasar and the galaxy group. We interpret these streams to be gas stripping from the satellite galaxies likely due to ram pressure stripping of material through interaction the galaxy’s ISM with the hot gas produced through quasar driven outflows.
\end{enumerate}
\textbf{Data availability}\\
The Keck OSIRIS data of this work are publicly available from the Keck Observatory Archive (https://www2.keck.hawaii.edu/koa/public/koa.php). Source information is provided with this paper. Other data underlying this article will be shared on a reasonable request to the corresponding author.
\\
\textbf{Acknowledgments}
The authors wish to thanks Jim Lyke, Randy Campbell, and other SAs with their assistance at the telescope to acquire the Keck OSIRIS data sets. We would like to thank Erica Keller, Melissa Hoffman, and Loreto Barcos Munoz for assistance with ALMA data reduction and imaging at NRAO. This paper makes use of the following ALMA data: ADS/JAO.ALMA 2013.1.01359.S, ADS/JAO.ALMA 2017.1.01527.S. ALMA is a partnership of ESO (representing its member states), NSF (USA) and NINS (Japan), together with NRC (Canada), MOST and ASIAA (Taiwan), and KASI (Republic of Korea), in cooperation with the Republic of Chile. The Joint ALMA Observatory is operated by ESO, AUI/NRAO and NAOJ. The National Radio Astronomy Observatory is a facility of the National Science Foundation operated under cooperative agreement by Associated Universities, Inc. The data presented herein were obtained at the W.M. Keck Observatory, which is operated as a scientific partnership among the California Institute of Technology, the University of California and the National Aeronautics and Space Administration. The Observatory was made possible by the generous financial support of the W.M. Keck Foundation. The authors wish to recognize and acknowledge the very significant cultural role and reverence that the summit of Maunakea has always had within the indigenous Hawaiian community. We are most fortunate to have the opportunity to conduct observations from this mountain. This research has made use of the NASA/IPAC Extragalactic Database (NED) which is operated by the Jet Propulsion Laboratory, California Institute of Technology, under contract with the National Aeronautics and Space Administration.

A.V., N.L.Z. and Y.I. acknowledge support from NASA ADAP grant 80NSSC21K1569. N.L.Z. was supported at the IAS by the J. Robert Oppenheimer Visiting Professorship and the Bershadsky Fund. We want to thank the anonymous referee for their constructive comments that helped improve the manuscript.

A.V. and N.L.Z. would like to thank Wuji Wang and Dominika Wylezalek for excellent discussions about CGM science with resolved \ly\ absorption in halos of luminous quasars.

\pagebreak
\bibliography{bib, nadia}{}
\bibliographystyle{aasjournal}

\appendix
\section{Galaxy group spectra}
\label{sec:appendix_sat}
\begin{figure*}[!th]
    \centering
    \includegraphics[height=4.5 in]{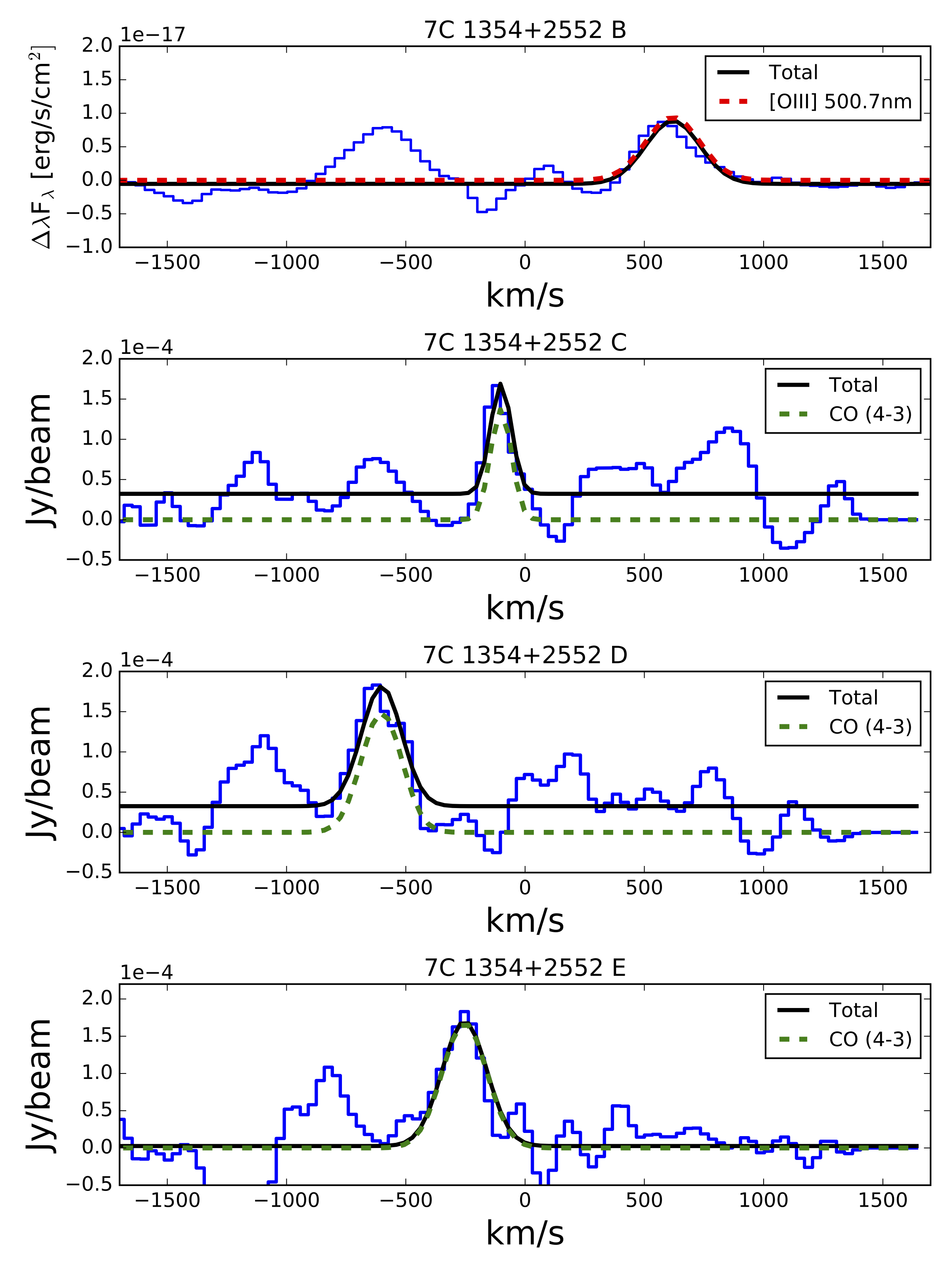}
    \includegraphics[height=4.5 in]{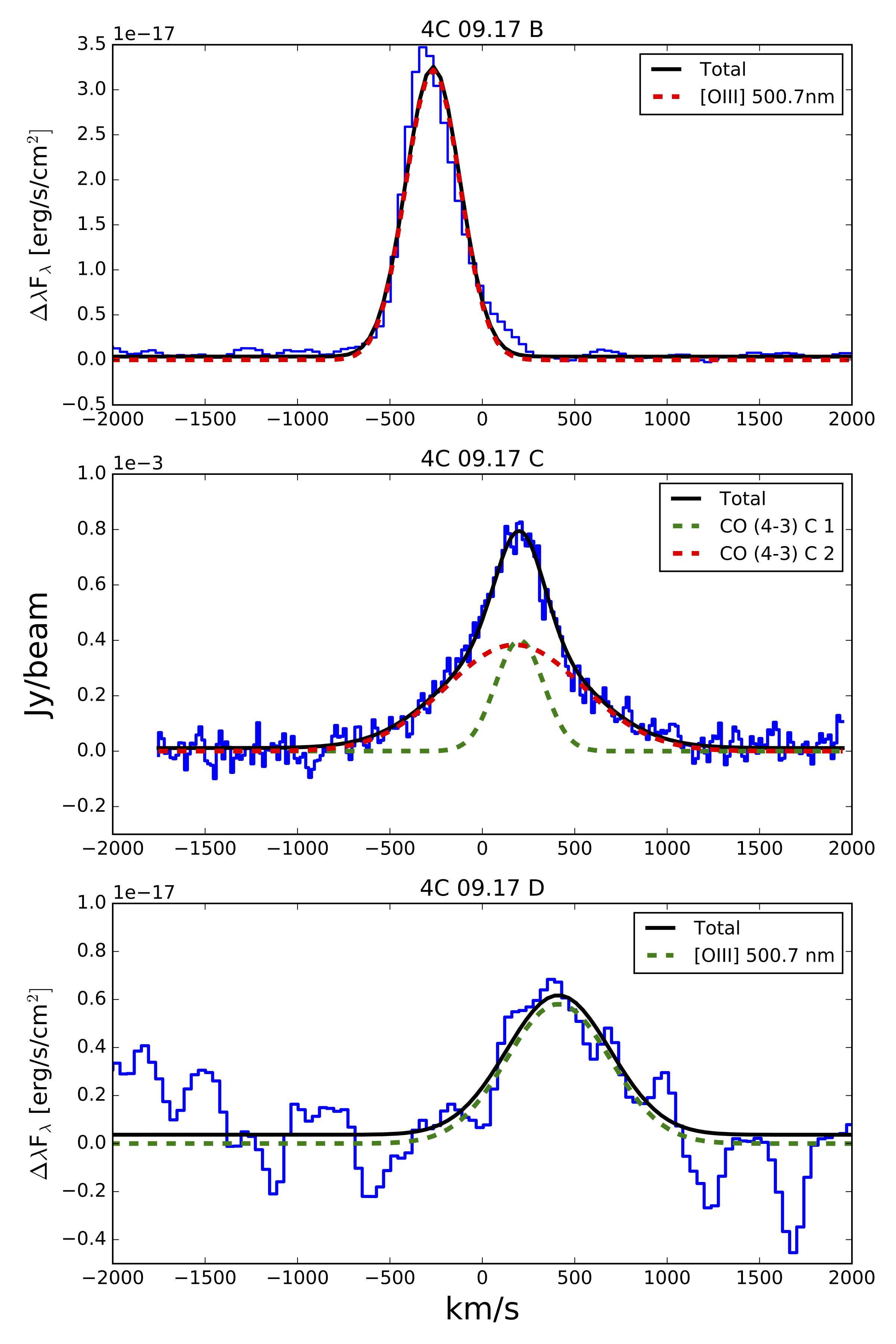}
    \caption{Spectra of individual galaxies in the group surrounding the 7C 1354+2552 quasar (left) and those surrounding the 4C 09.17 quasar (right). These spectra were used to identify and measure the redshift and velocity offset of each companion galaxy. We present the Gaussian model fits to each emission line.}
    \label{fig:companion_spectra}
\end{figure*}

\section{Emission and absorption line fitting}
\label{sec:appendix_fit}

In this section, we describe how we perform the emission and absorption line fitting for larger distinct regions in the individual sources. The \ly, \civ, and \heii\ emission lines are all fit with a combination of Gaussian profiles. The free parameters are the amplitude of the Gaussian profile, wavelength offset ($\lambda_{0}$) in the observed frame, and the velocity dispersion ($\sigma_{\lambda}$). For absorption we use an exponential profile: $\exp(-\tau(\lambda))$, where $\tau(\lambda)$ is the optical depth as a function of observed wavelength of the following form:

\begin{equation}
    \tau(\lambda) = \frac{\sqrt{\pi}e^{2}f_{i}\lambda_{0}^{2}}{\Delta \lambda_{D}m_{e}c^{2}}\times N_{i}\times H(a,x(\lambda))
\end{equation}
\noindent
$f_{i}$ is the oscillator strength, $e$ is the electron charge, $\Delta \lambda_{D}$ is defined as $b/c\times \lambda_{0}$ where $b$ is the thermal or Doppler broadening parameter, $m_{e}$ is the electron mass, $c$ is the speed of light, $N_{i}$ is the column density and $H(a,x(\lambda))$ is the Voigt-Hjerting function used to describe the shape of the absorption profile. The Voigt-Hjerting function is defined as:

\begin{equation}
    H(a,x(\lambda)) = \frac{a}{\pi} \int_{-\infty}^{+\infty} \frac{{e}^{-y^{2}}}{(x-y)^{2}+a^{2}} dy
\end{equation}

\noindent
where $x = (\lambda - \lambda_{0})/\Delta \lambda_{D}$ and $y=v/b$. The parameter a is defined as following:
\begin{equation*}
    a = \frac{\lambda_{0}A_{i}}{4\pi c \Delta \lambda_{D}}
\end{equation*}
\noindent
where $A_{i}$ is the Einstein A-coefficient. For $H(a,x(\lambda)$ we use the analytic approximation from \citet{TG06,TG07}: 

\begin{equation}
    H(a,x(\lambda)) = H_{0}-\frac{a}{\sqrt{\pi}x}\times(H_{0}\times H_{0}(4x^{4}+7x^{2}+4+Q)-Q-1)
\end{equation}

\noindent where $H_{0}=\exp(-x^{2})$ and $Q=1.5x^{-2}$. All atomic data is taken from physics.nist.gov. The final function that we fit the data is of the following form:

\begin{equation*}
    F(\lambda)=\sum_{n=1}^{k} F_{G,\lambda}^{n} \times \exp\left({-\sum_{j=1}^{l} \tau_{j,\lambda}}\right)
\end{equation*}

\noindent convolved with the line-spread function of KCWI before the fitting process begins. We first fit the data using a Least-Squares algorithm from Scipy. We then follow up with an MCMC routing using the \textit{emcee} \citep{Foreman-Mackey13} package. We use the best-fit parameters from the Least-Squares fit as the starting point for each walker, with a minor perturbation. First, we initialize 1000 walkers for each free parameter. We then run MCMC for 500 steps starting from the perturbed initial value. The priors on the free parameters are listed in Table \ref{tab:MCMC_range}. Before extracting the best fit parameters we discard 50 steps from the final chain.

\begin{table}[]
    \centering
    \caption{Table presenting the range of the priors for the free parameters for the Least-Squares and MCMC fitting algorithms.}
    \begin{tabular}{c|c}
    
        Parameter & Prior range  \\
        \hline
        Gaussian & \\
        \hline
        Amplitude & 0 - max(SNR) \\
        Offset ($\lambda_{0}$) & Location (max(SNR)) $\pm$ 4 \AA\\
        $\sigma_{v}$ & 40-4,000 \kms\\
        \hline
        Exponential absorption & \\
        \hline
        Offset ($\lambda_{0}$) & Location (min(SNR)) $\pm$ 2 \AA\\
        Doppler b & 40-400 \kms \\
        $N_{HI}$ & $10^{13}-10^{20}$cm$^{2}$ 
    \end{tabular}
    \label{tab:MCMC_range}
\end{table}

\subsection{Spatial binning and fitting spatially resolved absorption lines.}
To fit the absorption lines in \ly\ across the spatial extent of the \ly\ halo, we needed to bin up the data spatially to increase the SNR. We chose to use Voronoi binning method to spatially bin the data at the smallest loss of spatial resolution and information. We use the Python based Voronoi binning code by \citet{Cappellari09}. We chose to perform the binning on the \ly\ moment 0 map of both kinematic components such that each hexagonal tessellation achieves an integrated SNR over the \ly\ line of at least 50. We found this SNR to be optimal at fitting multiple absorption lines across the \ly\ line. We loop over the bins created by the Voronoi binning and extract the spectrum of each bin by averaging the data cube spatially at each spectral location. We then perform the exact same absorption and emission line fitting described above for the integrated spectrum of each individual kinematic component. From the best fit parameters we construct resolved radial velocity and equivalent width maps for each absorber.

\section{Morphology and extent of additional lines detected in the halos around 4C09.17 and 7C1354+2552.}
\label{sec:appendix_additional_lines}
\begin{figure*}
    \centering
    \includegraphics[width=5.5in]{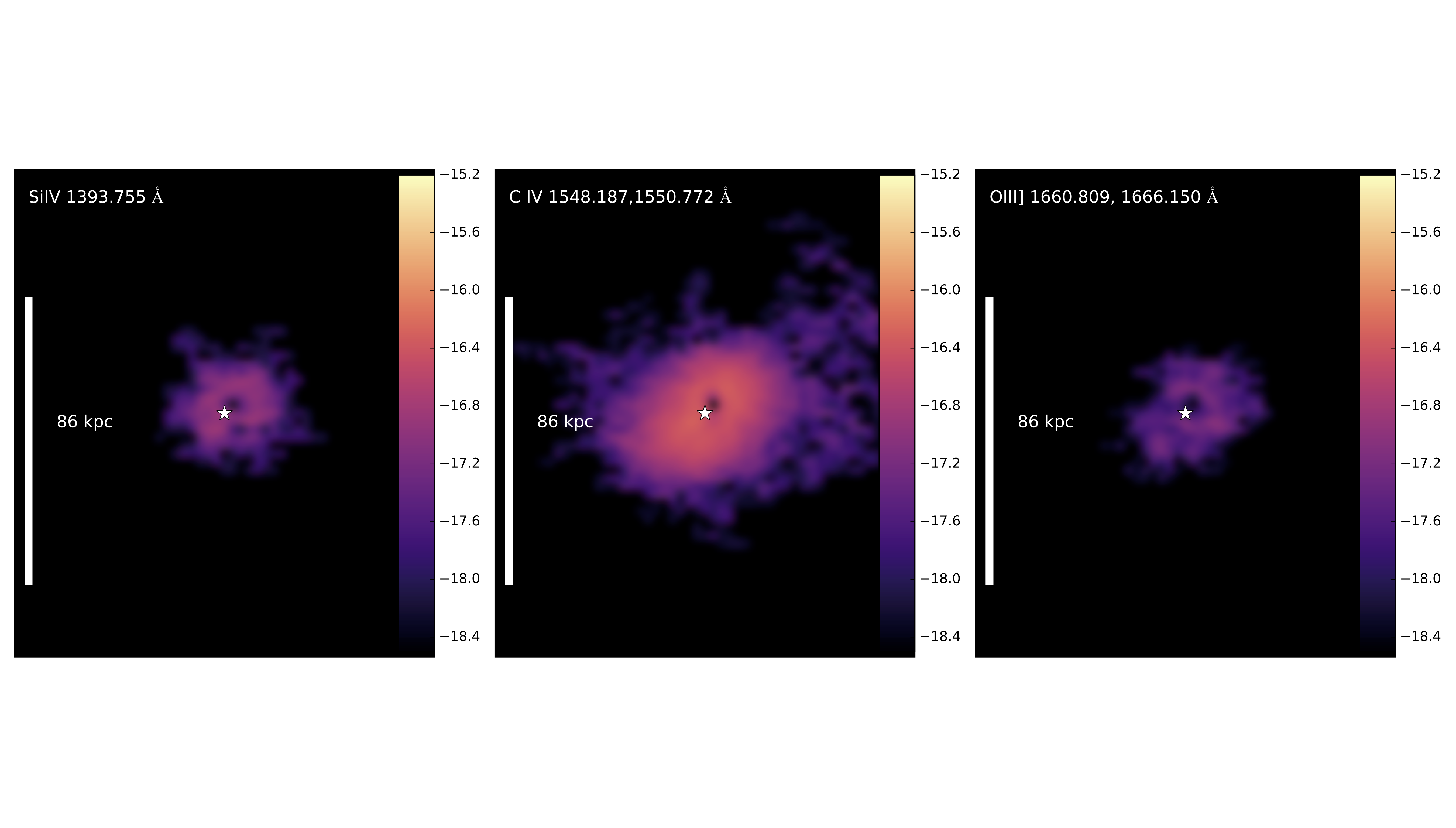}\\
    \includegraphics[width=5.5in]{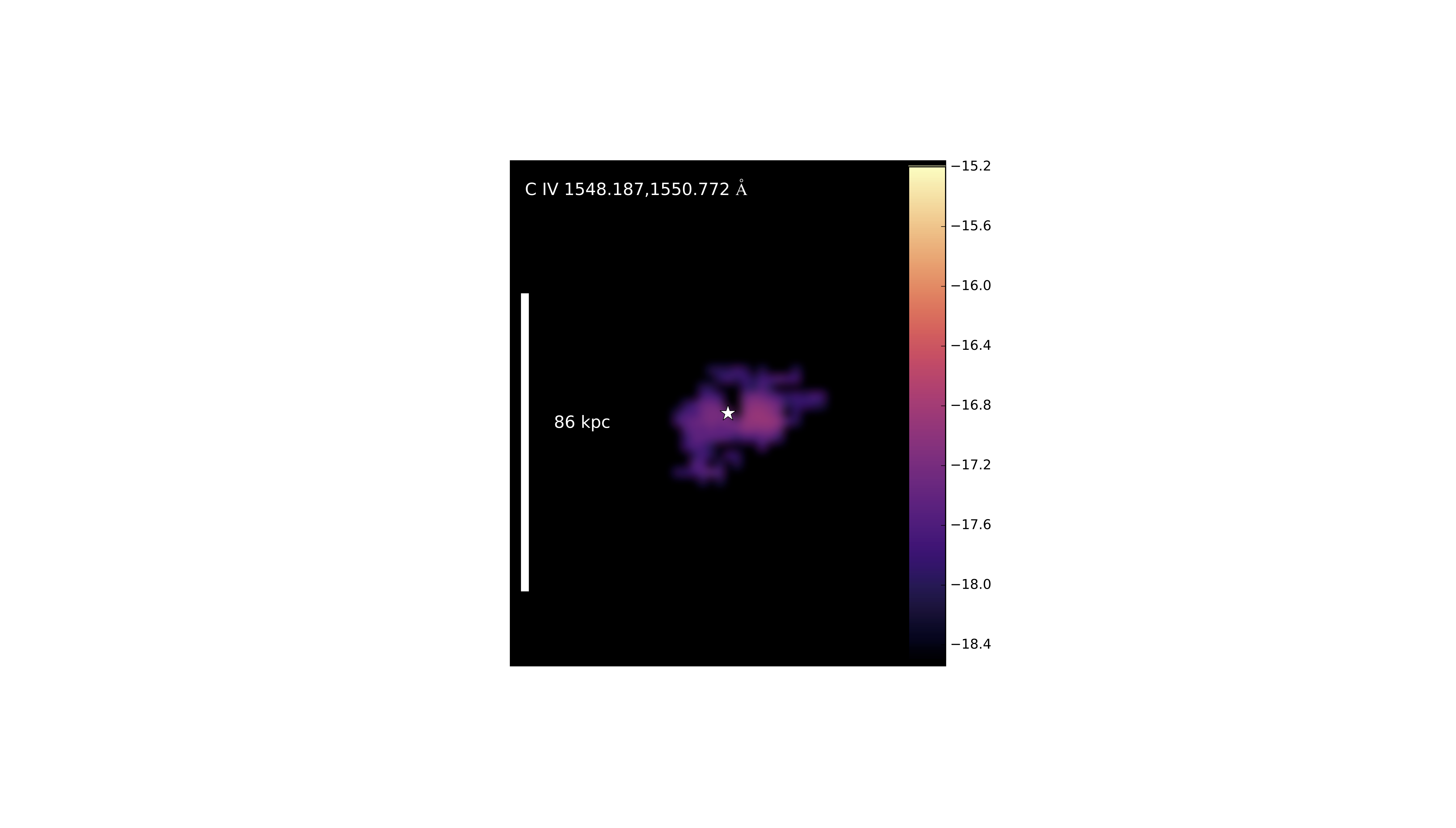}
    \caption{Surface brightness maps of additional fainter emission lines detected in both system, top: 7C 1354+2552, bottom: 4C 09.17. The star represents the location of the quasar and the bar to the left represents 10\arcsec\ or approximately 86 kpc at the redshift of our sources.}
    \label{fig:additional_lines}
\end{figure*}

\end{document}

%% file: macros.tex
\newcommand{\etal}{et al.}
\newcommand{\eg}{see, e.g.,}
\newcommand{\msun}{M$_{\sun}$\,}
\newcommand{\myr}{M$_\odot$~yr$^{-1}$} 
\newcommand{\ha}{H$\alpha$}
\newcommand{\hb}{H$\beta$}
\newcommand{\nii}{[N\textrm{\scshape{ii}}] }
\newcommand{\oiii}{[O\textrm{\scshape{iii}}] } 
\newcommand{\oii}{[O\textrm{\scshape{ii}}] } 
\newcommand{\oi}{[O\textrm{\scshape{i}}] } 
\newcommand{\hii}{H\textrm{\scshape{ii}}} 
\newcommand{\hi}{H\textrm{\scshape{i}}} 
\newcommand{\ly}{$\rm Ly\alpha$}
\newcommand{\heii}{$\rm HeII$}
\newcommand{\civ}{CIV}
\newcommand{\kms}{km\,s$^{-1}$} 
\newcommand{\rchi}{$\tilde{\chi}^{2}$}
\newcommand{\eden}{cm$^{-3}$}
\newcommand{\ergs}{erg s$^{-1}$}

\def\us{\char`\_}

\def\procspie{Proc. SPIE}
\def\newar{NewAR}
\def\pasj{PASJ}
\def\newblock{\hskip .11em plus .33em minus .07em}
\newcommand\ts{\thinspace}

\def\sun{\hbox{$\odot$}}
\def\earth{\hbox{$\oplus$}}
\def\lesssim{\mathrel{\hbox{\rlap{\hbox{\lower4pt\hbox{$\sim$}}}\hbox{$<$}}}}
\def\gtrsim{\mathrel{\hbox{\rlap{\hbox{\lower4pt\hbox{$\sim$}}}\hbox{$>$}}}}
\def\sq{\hbox{\rlap{$\sqcap$}$\sqcup$}}
\def\arcdeg{\hbox{$^\circ$}}
\def\arcmin{\hbox{$^\prime$}}
\def\arcsec{\hbox{$^{\prime\prime}$}}
\def\fd{\hbox{$.\!\!^{\rm d}$}}
\def\fh{\hbox{$.\!\!^{\rm h}$}}
\def\fm{\hbox{$.\!\!^{\rm m}$}}
\def\fs{\hbox{$.\!\!^{\rm s}$}}
\def\fdg{\hbox{$.\!\!^\circ$}}
\def\farcm{\hbox{.\kern -0.7ex\raisebox{.9ex}{\scriptsize$\prime$}}}
\def\farcs{\hbox{\kern 0.13ex.\kern -0.95ex%
\raisebox{.9ex}{\scriptsize$\prime\prime$}\kern -0.1ex}}
\def\fp{\hbox{$.\!\!^{\scriptscriptstyle\rm p}$}}
\def\micron{\hbox{$\mu$m}}

\newcommand{\brg}{Br$\gamma$}
\newcommand{\obs}{\it data}
\newcommand{\nsf}{NSFCAM}
\newcommand{\osc}{OSCIR}
\newcommand{\lw}{LWS}
\newcommand{\sig}{$\pm$}
\newcommand{\ulim}{$>$}
\newcommand{\kl}{$K-L$}
\newcommand{\kn}{$K-N$}
\newcommand{\knobs}{$K-[N]$}
\newcommand{\lncolor}{$L-N$}
\newcommand{\sigm}{$\sigma$}
\newcommand{\app}{$\sim$}

\newcommand*{\rom}[1]{\expandafter\@slowromancap\romannumeral #1@}

\newcommand{\myrkpc}{M$_\odot$~yr$^{-1}$~kpc$^{-2}$} 
\newcommand{\av}{A$_{v}$}

\newcommand{\sii}{[S{\sc II}]}

\newcommand{\loghn}{log([N\textrm{\scshape{ii}}]/H$\alpha$)}
\newcommand{\logohb}{log(\oiii/\hb) }
\newcommand{\logsh}{log(\sii/\ha) }
\newcommand{\logho}{log(\oi/\ha) }

\newcommand{\ferg}{erg s$^{-1}$ cm$^{-2}$}
\newcommand{\mjy}{mJy}
\newcommand{\htwo}{H$_{2}$}
\newcommand{\momfluxsfr}{$\dot{P}_{SFR}$ }
\newcommand{\momfluxagn}{$\dot{P}_{QSO}$ }
\newcommand{\momfluxout}{$\dot{P}_{outflow}$ }
\newcommand{\kineticlum}{$\dot{E}_{outflow}$ }
\newcommand{\energydepsne}{$\dot{E}_{SNe}$ }
\newcommand{\vsigmastellar}{$\sigma_{*}$ }
\newcommand{\vsigmagas}{$\sigma_{gas}$ }
\newcommand{\momfluxratio}{$\frac{\dot{P}_{outflow}}{\dot{P}_{AGN}}$}
\newcommand{\kineticlumratio}{$\frac{\dot{E}_{outflow}}{L_{Bol}}$}
\newcommand{\lbolqso}{$L_{Bol}$}
\newcommand{\msigma}{$M_{\bullet}-\sigma~$}
\newcommand{\mstellar}{$M_{\bullet}-M_{*}~$}

\def\sun{\hbox{$\odot$}}
\def\apj{ApJ}
\def\apjl{ApJL}
\def\aj{AJ}
\def\nat{Nature}
\def\araa{ARAA}
\def\aaps{A\&AS}
\def\apjs{ApJS}
\def\it{}
\def\mnras{MNRAS}
\def\aap{AAP}
\def\pasp{PASP}
\def\prd{Phys. Rev. D}
\def\earth{\hbox{$\oplus$}}
\def\lesssim{\mathrel{\hbox{\rlap{\hbox{\lower4pt\hbox{$\sim$}}}\hbox{$<$}}}}
\def\gtrsim{\mathrel{\hbox{\rlap{\hbox{\lower4pt\hbox{$\sim$}}}\hbox{$>$}}}}
\def\sq{\hbox{\rlap{$\sqcap$}$\sqcup$}}
\def\arcdeg{\hbox{$^\circ$}}
\def\arcmin{\hbox{$^\prime$}}
\def\arcsec{\hbox{$^{\prime\prime}$}}
\def\fd{\hbox{$.\!\!^{\rm d}$}}
\def\fh{\hbox{$.\!\!^{\rm h}$}}
\def\fm{\hbox{$.\!\!^{\rm m}$}}
\def\fs{\hbox{$.\!\!^{\rm s}$}}
\def\fdg{\hbox{$.\!\!^\circ$}}
\def\farcm{\hbox{.\kern -0.7ex\raisebox{.9ex}{\scriptsize$\prime$}}}
\def\farcs{\hbox{\kern 0.13ex.\kern -0.95ex%
\raisebox{.9ex}{\scriptsize$\prime\prime$}\kern -0.1ex}}
\def\fp{\hbox{$.\!\!^{\scriptscriptstyle\rm p}$}}

\def\ab{$\sim$}
\def\degr{^{\circ}}
\def\cd{CoD $-$33$\deg$7795 }
\def\Ha{H$\alpha$ }
\def\ha{H$\alpha$ }
\def\arcsec{''}
\def\sub{$\circ$}
\def\deg{\hbox{$^\circ$}}
\def\la{\mathrel{\hbox{\rlap{\hbox{\lower4pt\hbox{$\sim$}}}\hbox{$<$}}}}
\def\ga{\mathrel{\hbox{\rlap{\hbox{\lower4pt\hbox{$\sim$}}}\hbox{$>$}}}}
\def\sq{\hbox{\rlap{$\sqcap$}$\sqcup$}}
\def\arcmin{\hbox{$^\prime$}}
\def\arcsec{\hbox{$^{\prime\prime}$}}
\def\fd{\hbox{$.\!\!^{\rm d}$}}
\def\fh{\hbox{$.\!\!^{\rm h}$}}
\def\fm{\hbox{$.\!\!^{\rm m}$}}
\def\fs{\hbox{$.\!\!^{\rm s}$}}
\def\fdg{\hbox{$.\!\!^\circ$}}
\def\farcm{\hbox{$.\mkern-4mu^\prime$}}
\def\farcs{\hbox{$.\!\!^{\prime\prime}$}}
\def\fp{\hbox{$.\!\!^{\scriptscriptstyle\rm p}$}}
\def\onehalf{\hbox{$\,^1\!/_2$}}        
\def\onethird{\hbox{$\,^1\!/_3$}}
\def\twothirds{\hbox{$\,^2\!/_3$}}
\def\onequarter{\hbox{$\,^1\!/_4$}}
\def\threequarters{\hbox{$\,^3\!/_4$}}
\def\plotone{\includegraphics}